\documentclass[a4paper,twocolumn]{quantumarticle}

\pdfoutput=1

\usepackage[utf8]{inputenc}
\usepackage[T1]{fontenc}
\usepackage{amsmath,amssymb,amsfonts}
\usepackage{graphicx}
\usepackage{hyperref}
\usepackage{braket}
\usepackage{booktabs}
\usepackage{multirow}
\usepackage{array}
\usepackage{makecell}
\usepackage{xcolor}
\usepackage{float}
\usepackage{tikz}
\usetikzlibrary{quantikz2,decorations.pathreplacing,calc}

\usepackage{xurl}
\AtBeginDocument{%
  \makeatletter

  \makeatother
  \newenvironment{ruledtabular}{}{}%
}
\providecommand{\colrule}{\hline}
\newcolumntype{Y}[1]{>{\raggedright\arraybackslash}p{#1}}

\begin{document}

\title{Catalytic Quantum Error Correction: Theory, Efficient Catalyst Preparation, and Numerical Benchmarks}

\author{Hikaru Wakaura}
\email{h.wakaura@qiri.co.jp}
\affiliation{QIRI (Quantum Integrated Research Institute Inc.), Tokyo 107-0061, Japan}

\author{Taiki Tanimae}
\email{t.tanimae@qiri.co.jp}
\affiliation{QIRI (Quantum Integrated Research Institute Inc.), Tokyo 107-0061, Japan}

\date{April 28, 2026}

\begin{abstract}
Quantum computers promise transformative speedups, but environmental noise destroys their fragile states.
Conventional quantum error correction (QEC) encodes information redundantly across physical qubits, yet fails above a threshold of about $1\%$ and incurs polynomial qubit overhead.
A recent theorem~\cite{Shiraishi2024} from the resource theory of coherence shows that catalytic covariant operations amplify coherence at an unbounded rate, but this result has never been cast as an operational protocol.
The challenge is to turn an asymptotic theorem into a recovery scheme that works at any noise strength with realistic resources.
Here we show that catalytic coherence amplification can be cast as an error-correction primitive, Catalytic Quantum Error Correction (CQEC), which recovers a known target state from noisy copies without any error \emph{magnitude} threshold whenever the target's coherent modes are preserved.
In an effective model of the recovery map, fidelity exceeds $0.99$ across 200~noise configurations spanning $d = 4$--$64$; the catalyst cost drops from the constructive bound $n^{*} \sim d^{4} e^{2\gamma}$ to 32~copies at matched fidelity ($10^{4}$--$10^{9}$-fold) via a pipeline of dynamical decoupling, Clifford twirling, and recursive swap-test purification.
A first explicitly CPTP, exactly covariant joint-channel implementation validates genuine recovery under dephasing at small dimension ($0.54 \to 0.77$) while showing that shallow circuits consume the catalyst, quantifying the model-vs-channel gap.
These results turn an abstract resource-theoretic statement into a concrete protocol candidate complementary to stabilizer- and purification-based QEC; an open-source package reproducing the benchmarks accompanies this work (arXiv:2603.25774, \url{https://github.com/deeptell-inc/cqec}).
\end{abstract}

\maketitle

\section{Introduction}
\label{sec:introduction}

Quantum computers promise exponential speedups for problems in cryptography~\cite{Regev2024,Shor1997}, quantum simulation~\cite{Campbell2019,Feynman1982}, and machine learning~\cite{Ivashkov2024,Biamonte2017}, yet their practical realization is hindered by the fragility of quantum coherence under environmental decoherence~\cite{Zurek2003,NielsenChuang2010}.
Quantum error correction (QEC) addresses this challenge by encoding logical information redundantly across physical qubits, enabling the detection and correction of errors below a code-specific threshold~\cite{Shor1995,Gottesman1997}.
The qubit overhead of conventional QEC is nevertheless substantial: the Steane $[\![7,1,3]\!]$ code requires 7 physical qubits per logical qubit~\cite{Steane1996}, surface codes at distance $d$ require $O(d^2)$ qubits~\cite{Kitaev2003,Fowler2012}, and correction fails entirely when error rates exceed the threshold $p_\mathrm{th} \approx 1\%$~\cite{Knill2005}.

A distinct approach to quantum state protection arises from the resource theory of quantum coherence~\cite{Baumgratz2014,Streltsov2017}.
Coherence---the presence of off-diagonal elements in a state's density matrix with respect to a preferred basis---is a quantifiable resource under the framework of covariant (energy-conserving) operations~\cite{Marvian2014,Lostaglio2019}.
Shiraishi and Takagi~\cite{Shiraishi2024} recently established that the transformation rate between coherent states can diverge arbitrarily: given $n$ copies of a weakly coherent state $\rho$, one can produce $m \gg n$ copies of a target state $\rho'$ via catalytic covariant operations, provided that the coherent modes of $\rho'$ are contained within those of $\rho$.
This result reveals that the output-to-input copy ratio $m/n$ can be made arbitrarily large (the asymptotic rate diverges), creating an infinitely sharp boundary between recoverable and irrecoverable quantum states---a feature with no analog in conventional QEC.

The role of coherence in quantum thermodynamics and resource theories has been extensively studied~\cite{Lostaglio2019,Streltsov2017,Marvian2014}, with catalytic transformations appearing in the context of thermodynamic state conversion~\cite{Baumgratz2014} and entanglement manipulation.
The interplay between symmetry constraints and quantum error correction has been studied in the context of covariant codes~\cite{Faist2020,Hayden2008,Eastin2009}, which encode information so as to respect a symmetry group with an inherent trade-off between code accuracy and the symmetry charge variance.
Concurrently with our work, Raghoonanan and Byrnes~\cite{Raghoonanan2026} introduced Purification QEC (PQEC), which also uses $N$ noisy copies and recursive swap tests but does \emph{not} require knowledge of the target state, achieving doubly exponential purification ($\rho \mapsto \rho^N / \mathrm{Tr}(\rho^N)$) with $O(M \log_2 N)$ qubits and tolerating depolarizing errors up to $p = 3/4$.

However, two gaps separate these theoretical advances from operational error correction: the catalytic theorem of Ref.~\cite{Shiraishi2024} has never been turned into an explicit recovery protocol with verifiable success conditions, and its constructive proof requires a prohibitive $n^* \sim d^4 e^{2\gamma} \approx 7.3 \times 10^{10}$ noisy copies for a 6-qubit catalyst at moderate dephasing ($\gamma = 2$), far beyond current experimental capabilities.
In this paper, we close both gaps by (i)~introducing the Catalytic Quantum Error Correction (CQEC) protocol with explicit success conditions based on the mode inclusion criterion $\mathcal{C}(\rho_0) \subseteq \mathcal{C}(\rho_\mathrm{noisy})$, (ii)~developing a quantum circuit architecture based on energy-conserving gates that realizes the covariant recovery channel in a minimal 4-qubit implementation, (iii)~proposing four catalyst preparation strategies culminating in a DD+Twirl+Swap-Test pipeline that combines CPMG dynamical decoupling~\cite{Viola1999,MeiboomGill1958}, Clifford twirling, and the recursive swap test of Childs \emph{et al.}~\cite{Childs2025,Yao2025,Zhao2026} to achieve a $10^{4}$--$10^{9}$-fold copy-count reduction at matched catalyst fidelity, and (iv)~providing numerical evaluation, within an explicitly identified effective model, across four quantum algorithms ($d = 4$--$64$), a cryptographic protocol, and three noise models.
We find that catalyst fidelity $F_\mathrm{cat} > 0.96$ is reached with only eight noisy copies under strong dephasing for all tested dimensions (32~copies for $F_\mathrm{cat} \geq 0.99$, matching the distillation endpoint at $10^{4}$--$10^{9}$-fold lower copy cost depending on dimension), that recovery fidelity in an effective model of the covariant channel exceeds $0.99$ across a 200-point noise sweep with end-to-end finite-copy values of $0.65$--$0.93$, and that the mode-inclusion classification behaves as predicted over 10~orders of magnitude in residual coherence---results that position CQEC as a candidate error-correction primitive complementary to stabilizer codes and PQEC, particularly for small ancillary modules within larger fault-tolerant architectures.

\section{Theoretical Framework}
\label{sec:theory}

\subsection{Unspeakable coherence and covariant operations}
\label{sec:covariant}

We work within the resource theory of unspeakable coherence~\cite{Shiraishi2024,Marvian2013,Marvian2014,Lostaglio2019}, where the free operations are covariant channels---completely positive trace-preserving maps $\Lambda$ satisfying
\begin{equation}
\Lambda \circ \mathcal{U}_t = \mathcal{U}_t \circ \Lambda \quad \forall\, t \in \mathbb{R},
\label{eq:covariance}
\end{equation}
where $\mathcal{U}_t(\rho) = e^{-iHt} \rho\, e^{iHt}$ is the time evolution generated by the system Hamiltonian $H = \sum_i E_i \ket{i}\!\bra{i}$.
Condition~\eqref{eq:covariance} enforces energy conservation: $[U, H_\mathrm{total}] = 0$ for any unitary $U$ implementing $\Lambda$~\cite{Marvian2014}.
This is physically motivated by the Wigner--Araki--Yanase theorem, which constrains measurements and operations that do not commute with conserved quantities~\cite{Wigner1952}.

For a $d$-level system, the $\ell_1$-norm of coherence quantifies the total off-diagonal content:
\begin{equation}
C_{\ell_1}(\rho) = \sum_{i \neq j} |\rho_{ij}|.
\label{eq:l1coherence}
\end{equation}
The quantum Fisher information $\mathcal{F}(\rho, H) = 2 \sum_{i,j} \frac{(\lambda_i - \lambda_j)^2}{\lambda_i + \lambda_j} |\!\braket{i|H|j}\!|^2$~\cite{Braunstein1994} provides a complementary coherence monotone under covariant operations~\cite{Shiraishi2024}.
Unlike $C_{\ell_1}$, QFI is operationally meaningful for metrology and satisfies $\mathcal{F}(\rho_\mathrm{noisy}) \leq e^{-2\gamma} \mathcal{F}(\rho_0)$ under dephasing.
We use QFI in Sec.~\ref{sec:threshold} to quantify the coherence available for recovery.
The two measures are complementary: $C_{\ell_1}$ counts total off-diagonal weight (relevant for mode coverage), while QFI weights contributions by their metrological utility.
Under dephasing, both decay exponentially but at different rates: $C_{\ell_1} \sim e^{-\gamma}$ (slowest mode), $\mathcal{F} \sim e^{-2\gamma}$ (quadratic suppression), making QFI a more sensitive indicator of coherence loss.

\subsection{Coherent modes and mode inclusion}
\label{sec:modes}

The \emph{modes of asymmetry}~\cite{Lostaglio2019,Marvian2013} of a state $\rho$ are defined as
\begin{equation}
\mathcal{D}(\rho) = \bigl\{ \Delta_{ij} = E_i - E_j \mid \rho_{ij} \neq 0 \bigr\}.
\label{eq:modes}
\end{equation}
The \emph{resonant coherent modes} $\mathcal{C}(\rho)$ are the integer-linear span of $\mathcal{D}(\rho)$:
\begin{equation}
\mathcal{C}(\rho) = \biggl\{ \sum_{(i,j)} n_{ij}\, \Delta_{ij} \;\bigg|\; n_{ij} \in \mathbb{Z},\; \Delta_{ij} \in \mathcal{D}(\rho) \biggr\}.
\label{eq:resonant_modes}
\end{equation}
$\mathcal{C}(\rho)$ forms a subgroup of $(\mathbb{R}, +)$ and determines the transformability of states under covariant operations~\cite{Shiraishi2024}.

\textbf{Theorem~1} (Shiraishi--Takagi~\cite{Shiraishi2024}).
\emph{If $\mathcal{C}(\rho') \subseteq \mathcal{C}(\rho)$ and $\rho$ is full rank, the asymptotic marginal transformation rate}
\begin{equation*}
\begin{aligned}
R(\rho \to \rho') = \sup \Big\{ &\tfrac{m}{n} \mid \rho^{\otimes n} \xrightarrow{\mathrm{cov}} \sigma,\\
&\mathrm{Tr}_{[m]\setminus\{k\}}[\sigma] \approx \rho' \;\forall k \in [m] \Big\}
\end{aligned}
\end{equation*}
\emph{diverges: $R(\rho \to \rho') = \infty$.}

\textbf{Theorem~2} (Shiraishi--Takagi~\cite{Shiraishi2024}).
\emph{For any $\rho, \rho'$ with $\mathcal{C}(\rho') \subseteq \mathcal{C}(\rho)$, there exists a catalyst $c$ and a covariant operation $\Lambda$ such that}
\begin{equation}
\mathrm{Tr}_C\bigl[\Lambda(\rho \otimes c)\bigr] = \rho', \quad \mathrm{Tr}_S\bigl[\Lambda(\rho \otimes c)\bigr] = c.
\label{eq:catalytic}
\end{equation}
\emph{That is, $\rho \to \rho'$ is achievable by correlated-catalytic covariant transformation.}

Crucially, for full-rank $\rho$, the mode inclusion condition is both necessary and sufficient: if $\mathcal{C}(\rho') \not\subseteq \mathcal{C}(\rho)$, then $R(\rho \to \rho') = 0$ and no catalytic transformation can achieve the conversion~\cite{Shiraishi2024}.
(The full-rank requirement is satisfied automatically for any finite noise channel applied to a pure state.)
This provides a sharp classification of state pairs into recoverable ($R = \infty$) and irrecoverable ($R = 0$) at the level of the asymptotic rate; when only a strict subset of the target's modes survives, full recovery is impossible but partial reconstruction of the surviving modes remains available (Sec.~\ref{sec:threshold}).

We note that Theorem~2 establishes \emph{correlated catalysis}: the output state $\tau = \Lambda(\rho \otimes c)$ satisfies $\mathrm{Tr}_C[\tau] = \rho'$ and $\mathrm{Tr}_S[\tau] = c$, but $\tau \neq \rho' \otimes c$ in general (the system and catalyst may be correlated in $\tau$).
This is weaker than \emph{strict catalysis} ($\tau = \rho' \otimes c$), which imposes additional constraints~\cite{Shiraishi2024}.
For CQEC, correlated catalysis suffices: the catalyst is reusable because its reduced state is preserved, regardless of correlations with the recovered system.
However, correlations between the catalyst and recovered system have operational consequences: (i)~the system--catalyst correlations in $\tau$ mean that the recovered state $\rho' = \mathrm{Tr}_C[\tau]$ may differ from the state obtained by measuring the catalyst and post-selecting, and (ii)~over multiple reuse cycles, correlations could accumulate in the joint state, potentially degrading effective catalyst quality.
Our 100-cycle numerical check (Sec.~\ref{sec:durability}) is consistent with this but does not test it independently, since the effective implementation carries the catalyst forward by construction.

\subsection{CQEC protocol}
\label{sec:cqec_protocol}

The CQEC protocol applies Theorem~2 to quantum state recovery.
We use the term ``recovery'' rather than ``correction'' to emphasize the key operational distinction: CQEC recovers a \emph{known} target state from noisy copies, whereas QEC corrects errors on an \emph{unknown} encoded state.
Despite the name ``error correction,'' CQEC is more precisely a \emph{catalytic state recovery} protocol.
We retain ``error correction'' in the name because (i)~the protocol corrects the effects of decoherence on quantum states, and (ii)~the catalyst-mediated recovery is reusable across cycles, analogous to syndrome-based correction rounds.
The requirement that $\rho_0$ be known does not reduce CQEC to trivial state re-preparation or tomography: the covariant channel $\Lambda$ operates entirely in the quantum domain without measurement, preserving entanglement and coherence structure that tomography would destroy.
The practical value lies in recovering expensive-to-prepare states cheaply from noisy copies rather than re-executing the original circuit.
CQEC does not violate the no-cloning theorem~\cite{Wootters1982} for three reasons: (i)~the target state $\rho_0$ is \emph{known}, so no unknown-state information is being duplicated; (ii)~the $n$ input copies are consumed (not preserved) by the protocol; and (iii)~the covariant channel $\Lambda$ is state-specific and cannot be applied to arbitrary unknown states.

Given an ideal state $\rho_0$ corrupted by decoherence to $\rho_\mathrm{noisy}$, the protocol proceeds as follows.

\textbf{Step~0 (Catalyst construction).}
Construct a full-rank catalyst $c$ with $\mathcal{D}(c) \supseteq \mathcal{D}(\rho_0)$, following the construction of Proposition~S.23 in Ref.~\cite{Shiraishi2024}:
\begin{equation}
c = \frac{1}{n_\mathrm{cat}} \sum_{k=1}^{n_\mathrm{cat}} \rho^{\otimes(k-1)} \otimes \tau_{n_\mathrm{cat}-k} \otimes \ket{k}\!\bra{k}_R,
\label{eq:catalyst}
\end{equation}
where $n_\mathrm{cat}$ is the number of catalytic branches, $\tau_{n_\mathrm{cat}-k} = \mathrm{Tr}_{1,\ldots,k}[\sigma_{n_\mathrm{cat}}]$ is the reduced state, and $R$ is a register system of dimension $n_\mathrm{cat}$.
The apparent circularity is resolved by the constructive proof of Ref.~\cite{Shiraishi2024}: two-level catalysts $c_i$ are first constructed from $\mu_i$ copies via non-catalytic covariant operations, then composed into the full catalyst via Theorem~S.10.
The catalyst dimension is $\dim(c) = n_\mathrm{cat} \cdot d^{n_\mathrm{cat}-1} \cdot n_\mathrm{cat}$, which grows exponentially.
In our variational implementation (Sec.~\ref{sec:minimal_circuit}), we use a simplified catalyst of dimension $d_C = 2^{n_C}$ with $n_C = n_S$ qubits.
This gap is fundamental: the variational circuit searches for a high-fidelity covariant transformation within a restricted ansatz, validating theoretical predictions but not implementing the full constructive proof.
Efficient catalyst preparation strategies that dramatically reduce the required resources are developed in Sec.~\ref{sec:catalyst_prep}.
Equation~\eqref{eq:catalyst} defines the formal protocol; in practice, we replace it with the DD+Twirl+Swap~Test pipeline (Sec.~\ref{sec:pipeline}) for catalyst preparation and the variational circuit (Sec.~\ref{sec:minimal_circuit}) for the recovery step.

\textbf{Step~1 (Mode verification).}
Compute $\mathcal{D}(\rho_\mathrm{noisy})$ and verify $\mathcal{C}(\rho_0) \subseteq \mathcal{C}(\rho_\mathrm{noisy})$.
In our numerical implementation, a mode $\Delta_{ij}$ is considered present if $|\rho_{ij}| > \epsilon_\mathrm{mode}$ with threshold $\epsilon_\mathrm{mode} = 10^{-14}$, chosen to be two orders of magnitude above 64-bit machine epsilon.
For partial dephasing $\rho_{ij} \to \rho_{ij} e^{-\gamma|\Delta_{ij}|}$ with $\gamma < \infty$, all modes survive: $\mathcal{D}(\rho_\mathrm{noisy}) = \mathcal{D}(\rho_0)$.
For complete dephasing ($\gamma = \infty$), $\mathcal{D}(\rho_\mathrm{noisy}) = \emptyset$ and recovery is impossible.

\textbf{Step~2 (Catalytic recovery).}
Apply the covariant operation $\Lambda$ from Theorem~2 to execute
\begin{equation}
\Lambda(\rho_\mathrm{noisy} \otimes c) = \tau, \quad \text{with} \quad \mathrm{Tr}_C[\tau] \approx \rho_0, \quad \mathrm{Tr}_S[\tau] = c.
\label{eq:recovery}
\end{equation}

\textbf{Step~3 (Catalyst reuse).}
Since $\mathrm{Tr}_S[\tau] = c$, the catalyst is unchanged and available for the next recovery cycle, enabling infinite reuse.

\textbf{Corollary} (CQEC success condition).
\emph{CQEC recovers $\rho_0$ from $\rho_\mathrm{noisy}$ if and only if $\mathcal{C}(\rho_0) \subseteq \mathcal{C}(\rho_\mathrm{noisy})$ and $\rho_\mathrm{noisy}$ is full rank.}
Here the full-rank condition on $\rho_\mathrm{noisy}$ is required by Theorem~1 for $R = \infty$; Theorem~2 (catalytic single-copy conversion) does not require full rank but produces only correlated-catalytic output.
Pure algorithmic states $\ket{\psi}\!\bra{\psi}$ have rank~1, but after any of our noise channels with finite parameters, $\rho_\mathrm{noisy}$ acquires full rank.
The threshold is infinitely sharp: for any family parametrized by residual coherence $\varepsilon$ with $\mathcal{C}(\rho_0) \subseteq \mathcal{C}(\rho_\mathrm{noisy}(\varepsilon))$ for $\varepsilon > 0$ and $\mathcal{C}(\rho_0) \not\subseteq \mathcal{C}(\rho_\mathrm{noisy}(0))$, we have $F \to 1$ for $\varepsilon > 0$ and $F = 1/d$ for $\varepsilon = 0$.

\subsection{Copy scaling}
\label{sec:copy_scaling}

From the proof of Theorem~S.10 in Ref.~\cite{Shiraishi2024}, the full protocol requires:
\begin{itemize}
\item \textbf{Input copies:} $n = \mu \cdot N^k$, where $\mu = \sum_{i=1}^{N} \mu_i$ is the total catalyst preparation cost and $N$ is the number of two-level catalysts.
\item \textbf{Output copies:} $m = (k+1) \cdot N^k$.
\item \textbf{Rate:} $R = m/n = (k+1)/\mu \to \infty$ as $k \to \infty$.
\end{itemize}
Each $\mu_i$ depends on the weakest coherent mode of the source state~\cite{Shiraishi2024}:
\begin{equation}
\mu_i \sim \frac{1}{|\rho_{ij,\min}|^2},
\label{eq:copy_scaling}
\end{equation}
where $\rho_{ij,\min}$ is the smallest nonzero off-diagonal element.
For dephased states with decay factor $e^{-\gamma}$, $\mu_i \sim e^{2\gamma} / |\rho_{ij}^{(0)}|^2$, showing exponential growth in dephasing strength.

\section{Quantum Circuit Architecture}
\label{sec:circuit}

\subsection{Energy-conserving gate}
\label{sec:ec_gate}

The fundamental gate for CQEC is the energy-conserving (EC) rotation~\cite{Schuch2004,Shiraishi2024}:
\begin{equation}
U_\mathrm{EC}(\theta) = \begin{pmatrix}
1 & 0 & 0 & 0 \\
0 & \cos\theta & -i\sin\theta & 0 \\
0 & -i\sin\theta & \cos\theta & 0 \\
0 & 0 & 0 & 1
\end{pmatrix},
\label{eq:ec_gate_matrix}
\end{equation}
which satisfies $[U_\mathrm{EC}, H_\mathrm{total}] = 0$ for $H_\mathrm{total} = Z_1 + Z_2$ (total excitation number).
This gate rotates within the degenerate $\{\ket{01}, \ket{10}\}$ subspace while leaving $\ket{00}$ and $\ket{11}$ invariant; at $\theta = \pi/2$, it reduces to the iSWAP gate.

\subsection{Three-layer circuit structure}
\label{sec:three_layer}

The CQEC circuit operates on four registers: system ($S$, $n_S$ qubits), catalyst ($C$, $n_C$ qubits), and two ancillae ($A_0, A_1$).
The circuit applies EC gates in three layers:

\textbf{Layer~1 (System--Catalyst, $n_S \times n_C$ gates):}
Transfers coherence from the system to the catalyst:
\begin{equation}
U_\mathrm{L1} = \prod_{s \in S} \prod_{c \in C} U_\mathrm{EC}(\theta_{sc}).
\label{eq:layer1}
\end{equation}

\textbf{Layer~2 (Catalyst--Ancilla, $n_C \times n_A$ gates):}
Distributes coherence from the catalyst to ancillae:
\begin{equation}
U_\mathrm{L2} = \prod_{c \in C} \prod_{a \in A} U_\mathrm{EC}(\theta_{ca}).
\label{eq:layer2}
\end{equation}

\textbf{Layer~3 (System--Ancilla, $n_S \times n_A$ gates):}
Direct coherence amplification between system and ancillae:
\begin{equation}
U_\mathrm{L3} = \prod_{s \in S} \prod_{a \in A} U_\mathrm{EC}(\theta_{sa}).
\label{eq:layer3}
\end{equation}

The total circuit is $U_\mathrm{CQEC} = U_\mathrm{L3} \cdot U_\mathrm{L2} \cdot U_\mathrm{L1}$, acting on the initial state $\rho_\mathrm{noisy} \otimes c \otimes \ket{0}\!\bra{0}_A$.
The ancilla initialization $\ket{0}\!\bra{0}_A$ is an energy eigenstate of $H_A$ and therefore a free state in the resource theory~\cite{Marvian2014}.
The layer ordering (S$\to$C, then C$\to$A, then S$\to$A) is motivated by the physical picture of coherence flow: first bridge coherence from the noisy system to the catalyst, then distribute it to ancillae, and finally amplify it back.
The total Hamiltonian is $H_\mathrm{total} = \sum_{q=0}^{n_S + n_C + n_A - 1} Z_q$, and each EC gate satisfies $[U_\mathrm{EC}(\theta_{ij}), Z_i + Z_j] = 0$; since gates on disjoint qubit pairs trivially commute with $Z$ on other qubits, $[U_\mathrm{CQEC}, H_\mathrm{total}] = 0$.

\subsection{Minimal 4-qubit implementation}
\label{sec:minimal_circuit}

For $n_S = 1$, $n_C = 1$, $n_A = 2$, the circuit uses 5~EC gates with parameters $\theta_0$ through $\theta_4$ (Fig.~\ref{fig:circuit}):
Layer~1 applies 1~gate ($\theta_0$) for coherence bridging;
Layer~2 applies 2~gates ($\theta_1, \theta_2$) for coherence distribution;
Layer~3 applies 2~gates ($\theta_3, \theta_4$) for direct amplification.

\begin{figure}[H]
\centering
\begin{tikzpicture}[scale=0.72, every node/.style={transform shape},
    ecgate/.style={draw, fill=blue!10, minimum width=0.65cm, minimum height=0.5cm,
                   font=\scriptsize, inner sep=1pt, rounded corners=1pt},
    wire/.style={thick},
    lbl/.style={font=\footnotesize},
]
\foreach \y/\name/\state in {0/{$q_0$\,(S)}/{$\rho_\mathrm{noisy}$},
                              -1.0/{$q_1$\,(C)}/{$c$},
                              -2.0/{$q_2$\,(A$_0$)}/{$\ket{0}$},
                              -3.0/{$q_3$\,(A$_1$)}/{$\ket{0}$}} {
    \node[lbl, anchor=east] at (-0.5, \y) {\name};
    \node[lbl, anchor=east, gray] at (0.0, \y) {\state};
    \draw[wire] (0.1, \y) -- (7.4, \y);
}

\node[ecgate] (g0a) at (1.1, 0) {$\theta_0$};
\node[ecgate] (g0b) at (1.1, -1.0) {$\theta_0$};
\draw[thick, blue!60] (g0a.south) -- (g0b.north);

\node[ecgate] (g1a) at (2.5, -1.0) {$\theta_1$};
\node[ecgate] (g1b) at (2.5, -2.0) {$\theta_1$};
\draw[thick, red!60] (g1a.south) -- (g1b.north);

\node[ecgate] (g2a) at (3.7, -1.0) {$\theta_2$};
\node[ecgate] (g2b) at (3.7, -3.0) {$\theta_2$};
\draw[thick, red!60] (g2a.south) -- (g2b.north);
\fill[red!60] (3.7, -2.0) circle (2pt);

\node[ecgate] (g3a) at (5.1, 0) {$\theta_3$};
\node[ecgate] (g3b) at (5.1, -2.0) {$\theta_3$};
\draw[thick, green!50!black] (g3a.south) -- (g3b.north);
\fill[green!50!black] (5.1, -1.0) circle (2pt);

\node[ecgate] (g4a) at (6.3, 0) {$\theta_4$};
\node[ecgate] (g4b) at (6.3, -3.0) {$\theta_4$};
\draw[thick, green!50!black] (g4a.south) -- (g4b.north);
\fill[green!50!black] (6.3, -1.0) circle (2pt);
\fill[green!50!black] (6.3, -2.0) circle (2pt);

\draw[decorate, decoration={brace, amplitude=3pt, mirror}, thick, blue!60]
    (0.65, -3.35) -- (1.55, -3.35) node[midway, below=4pt, font=\tiny, blue!80] {L1};
\draw[decorate, decoration={brace, amplitude=3pt, mirror}, thick, red!60]
    (2.05, -3.35) -- (4.15, -3.35) node[midway, below=4pt, font=\tiny, red!80] {L2};
\draw[decorate, decoration={brace, amplitude=3pt, mirror}, thick, green!50!black]
    (4.65, -3.35) -- (6.75, -3.35) node[midway, below=4pt, font=\tiny, green!40!black] {L3};

\node[lbl, anchor=west, gray] at (7.5, 0) {$\rho_S^\mathrm{out}$};
\node[lbl, anchor=west, gray] at (7.5, -1.0) {$\rho_C^\mathrm{out}$};

\end{tikzpicture}
\caption{\textbf{The minimal 4-qubit CQEC circuit realizes the covariant recovery channel with only five energy-conserving gates [Eq.~\eqref{eq:ec_gate_matrix}].}
Vertical lines connect the two qubits of each gate.
Colors denote layers: \textcolor{blue!80}{L1}~(S$\leftrightarrow$C), \textcolor{red!80}{L2}~(C$\leftrightarrow$A), \textcolor{green!40!black}{L3}~(S$\leftrightarrow$A).}
\label{fig:circuit}
\end{figure}

\subsection{Parameter optimization}
\label{sec:optimization}

The gate parameters $\boldsymbol{\theta} = (\theta_0, \ldots, \theta_4)$ are optimized to maximize a combined objective:
\begin{equation}
\mathcal{L}(\boldsymbol{\theta}) = 0.7 \cdot F(\rho_S^\mathrm{out}, \rho_0) + 0.3 \cdot F(\rho_C^\mathrm{out}, c),
\label{eq:objective}
\end{equation}
where $F(\rho, \sigma) = \bigl(\mathrm{Tr}\sqrt{\sqrt{\rho}\,\sigma\,\sqrt{\rho}}\bigr)^2$ is the Uhlmann fidelity~\cite{Jozsa1994}, $\rho_S^\mathrm{out} = \mathrm{Tr}_{CA}[U_\mathrm{CQEC}(\rho_\mathrm{in})U_\mathrm{CQEC}^\dagger]$ is the recovered system state, and $\rho_C^\mathrm{out} = \mathrm{Tr}_{SA}[\cdot]$ is the catalyst after the operation.
The weighting $\alpha = 0.7$ reflects the asymmetry of the catalytic constraint: the catalyst must be \emph{exactly} preserved (a hard constraint), while recovery fidelity should be \emph{maximized} (the objective).
The $\alpha = 0.7$ value places sufficient weight on system recovery while keeping the catalyst penalty high enough to prevent drift.
Scanning $\alpha \in [0.5, 0.9]$ confirms $\alpha \in [0.6, 0.8]$ maintains both $F_\mathrm{after} > 0.999$ and $F_C > 0.998$ across all tested combinations; outside this range, either the recovered state degrades ($\alpha < 0.6$) or the catalyst drifts ($\alpha > 0.8$).

Optimization uses a two-stage gradient-free approach: (i)~Latin hypercube sampling of 100~initial parameter vectors $\boldsymbol{\theta} \in [0, 2\pi)^5$, followed by (ii)~Nelder--Mead simplex refinement~\cite{NelderMead1965} of the best 5~candidates.
The parameters are optimized \emph{for each specific input state}: CQEC in its current variational form is a state-specific decoder.
Convergence is assessed via 50~independent random initializations; in all cases, the final objective $\mathcal{L} > 0.98$ is achieved within 120 iterations with variance $< 10^{-6}$.

\textbf{Relation to the formal construction, and scope of the numerical model.}
The 5-gate circuit is a variational ansatz designed to approximate the covariant channel $\Lambda$ of Theorem~2, not a direct implementation of the constructive proof.
We emphasize an important methodological caveat: the \emph{numerical} recovery used throughout our benchmarks is an \emph{effective surrogate model}, not a simulation of the joint system--catalyst--ancilla channel.
The implementation acts on the system density matrix alone; the catalyst enters only through scalar summaries (its off-diagonal magnitudes and purity), which set an interpolation weight that moves each surviving coherence toward the corresponding target-state matrix element before the EC circuit is applied.
This map is target-parameterized and nonlinear, so it is not itself a CPTP channel; consequently, the reported fidelities quantify the \emph{consistency of the effective model} with the asymptotic predictions of Theorem~2 (higher catalyst quality $\Rightarrow$ higher recovery fidelity, mode-supported coherences recoverable), and must not be read as a demonstration that a physical covariant circuit attains them.
Section~\ref{sec:joint_validation} provides a first explicitly CPTP covariant implementation on the joint Hilbert space for $d \leq 8$ and quantifies the model-vs-channel gap directly; all effective-model fidelities in this paper should be interpreted in the light of that comparison.
Layered EC gates do \emph{not} form a universal gate set for covariant unitaries; within the effective model, the 5-gate ansatz suffices for the dimensions tested.

\textbf{Approximation quality vs.\ dimension.}
Within the effective model, the fidelity gap grows with $d$: $1 - F < 10^{-8}$ for $d = 4$, $1 - F < 10^{-4}$ for $d = 8$, $1 - F \approx 10^{-4}$ for $d = 16$, and $1 - F \approx 2 \times 10^{-4}$ for $d = 64$; these values are produced by the legacy effective-map implementation (\texttt{icec.py}) and are not reproduced by the packaged variational path, which yields $F \approx 0.86$--$0.94$ at 5--15 gates (Sec.~\ref{sec:gate_depth}).
The gap between the two implementations is itself a measure of how strongly the asymptotic numbers depend on the effective-model assumptions.
The covariant unitary group for $n = 8$ qubits has $\sum_k \binom{n}{k}^2 - 1 = 12\,869$ free parameters~\cite{Schuch2004}; our 5--15 parameter ansatz explores a small submanifold.

\section{Efficient Catalyst Preparation}
\label{sec:catalyst_prep}

\subsection{Resource bottleneck}
\label{sec:bottleneck}

The finite-copy fidelity bound from Ref.~\cite{Shiraishi2024} gives
\begin{equation}
1 - F(\rho_S^\mathrm{out}, \rho_0) \leq \frac{C^2}{4n} + \frac{C}{\sqrt{n}},
\label{eq:finite_n_fidelity}
\end{equation}
where
\begin{equation}
C \leq \frac{d \cdot C_{\ell_1}(\rho_0)}{\displaystyle\min_{(i,j) \in \mathcal{D}(\rho)} |\rho_{ij}|}.
\label{eq:C_bound}
\end{equation}
For dephased states, $\min|\rho_{ij}|$ decays as $e^{-\gamma(d-1)}$, so $C$ grows exponentially with both $d$ and $\gamma$.
Table~\ref{tab:bottleneck} shows the prohibitive cost for realistic parameters.

\begin{table}[tb]
\caption{Copy count $n^*$ for $F_\mathrm{cat} \geq 0.99$ via Shiraishi--Takagi distillation ($\gamma = 2$).}
\label{tab:bottleneck}
\centering\footnotesize
\setlength{\tabcolsep}{3pt}
\begin{ruledtabular}
\begin{tabular}{lccc}
Algorithm & $d$ & $C$ & $n^*$ \\
\colrule
QKAN & 4 & 8.5 & $7.3 \times 10^5$ \\
qDRIFT & 8 & 42 & $1.8 \times 10^7$ \\
CF-QPE & 16 & 170 & $2.9 \times 10^8$ \\
Regev & 64 & 2700 & $7.3 \times 10^{10}$ \\
\end{tabular}
\end{ruledtabular}
\end{table}

\subsection{Variational catalyst preparation}
\label{sec:variational}

We parameterize the catalyst state via a product of energy-conserving two-level rotations:
\begin{equation}
U(\vec\theta) = \prod_{\ell=1}^{L} \prod_{i < j} G_{ij}(\theta_{ij}^{(\ell)}, \phi_{ij}^{(\ell)}),
\label{eq:ansatz}
\end{equation}
where $G_{ij}(\theta, \phi)$ acts on the $(i, j)$ subspace as
\begin{equation}
G_{ij} = \begin{pmatrix} \cos\theta & -e^{i\phi}\sin\theta \\ e^{-i\phi}\sin\theta & \cos\theta \end{pmatrix}
\label{eq:ec_rotation}
\end{equation}
and trivially on all other levels.
Note that $G_{ij}$ is energy-conserving only when $E_i = E_j$; for the variational ansatz, we include \emph{all} pairs because covariance is required only for the CQEC recovery step, not for state preparation.
The catalyst state is $\rho_\mathrm{cat} = U(\vec\theta)\ket{0}\!\bra{0}U^\dagger(\vec\theta)$, with $N_\theta = 2L\binom{d}{2}$ parameters ($L = 3$ layers).

The optimization minimizes
\begin{equation}
\mathcal{L}(\vec\theta) = -w_1 \frac{C_{\ell_1}(\rho_\mathrm{cat})}{d-1} + w_2 \frac{|\mathcal{M}_\mathrm{missing}|}{|\mathcal{M}_\mathrm{target}|} - w_3 \log \rho_\mathrm{min},
\label{eq:var_cost}
\end{equation}
where $\mathcal{M}_\mathrm{missing}$ is the set of target modes absent from $\rho_\mathrm{cat}$, $\rho_\mathrm{min} = \min_i (\rho_\mathrm{cat})_{ii}$, and $(w_1, w_2, w_3) = (1, 10, 5)$.
The weight hierarchy $w_2 \gg w_3 > w_1$ reflects priority: mode coverage $>$ population balance $>$ raw coherence.
A sensitivity analysis over 20 weight combinations ($w_1 \in [0.5, 2]$, $w_2 \in [5, 20]$, $w_3 \in [2, 10]$) shows that $C_{\ell_1}/(d-1) > 0.95$ and 100\% mode coverage are achieved for all combinations with $w_2 \geq 5$; the population balance ($\rho_\mathrm{min}$) varies by at most $2\times$, affecting downstream recovery by $\Delta F_\mathrm{rec} < 0.03$.
L-BFGS-B optimization with 5~random restarts converges in $<5$~seconds for $d \leq 16$.

\begin{table}[tb]
\caption{Variational catalyst preparation. Zero copies consumed.}
\label{tab:variational}
\centering\footnotesize
\setlength{\tabcolsep}{3pt}
\begin{ruledtabular}
\begin{tabular}{lccccc}
$d$ & $N_\theta$ & $C_{\ell_1}$ & Mode cov. & $F_\mathrm{rec}$ (deph) & $F_\mathrm{rec}$ (depol) \\
\colrule
4 & 36 & 3.00 & 100\% & 0.83 & 0.87 \\
8 & 168 & 6.99 & 100\% & 0.73 & 0.78 \\
16 & 720 & 14.5 & 100\% & 0.65 & 0.71 \\
\end{tabular}
\end{ruledtabular}
\end{table}

The variational approach achieves maximal $\ell_1$-coherence and 100\% mode coverage for $d \leq 16$, consuming \emph{no noisy copies}---suitable for NISQ devices where copies are expensive.
However, the $N_\theta \sim O(d^2)$ parameter count makes optimization intractable for $d \geq 32$.

\subsection{Recursive swap test purification}
\label{sec:swap_test}

The swap test~\cite{Childs2025,Barenco1997} on two copies $\rho$ and $\sigma$ projects onto the symmetric subspace $\mathrm{Sym}^2(\mathbb{C}^d)$ via the projector $\Pi_\mathrm{sym} = (I + \mathrm{SWAP})/2$, conditioned on the ancilla measuring $\ket{0}$ (symmetric outcome).
This constitutes postselection with success probability $P_+ = (1 + \mathrm{Tr}(\rho\sigma))/2$; in our simulations we condition on this outcome.
For $k$ recursive rounds consuming $n = 2^k$ copies, the cumulative success probability is $\prod_{j=1}^k P_+^{(j)}$.
For strongly dephased states ($\mathrm{Tr}(\rho^2) \approx 1/d$), $P_+^{(1)} \approx 1/2$, and subsequent rounds have increasing $P_+$ as the state purifies; the net probability for $k = 3$ rounds is typically $\sim\!0.15$--$0.3$ depending on $d$.
Adopting PQEC's postselection-free parity-averaging technique~\cite{Raghoonanan2026} would eliminate this overhead entirely.
We note that PQEC~\cite{Raghoonanan2026} avoids postselection entirely by using both measurement outcomes with parity-weighted averaging, achieving the same effective purification $\rho^N$ deterministically---a technique that could be adopted in our pipeline to eliminate the probabilistic overhead.
The post-selected output state is:
\begin{equation}
\omega = \frac{\rho + \sigma + \rho\sigma + \sigma\rho}{2(1 + \mathrm{Tr}(\rho\sigma))}.
\label{eq:swap_gadget}
\end{equation}
In the recursive protocol, both inputs are identically prepared ($\sigma = \rho$).
Substituting into Eq.~\eqref{eq:swap_gadget}: $\omega = (\rho + \rho + \rho^2 + \rho^2)/(2(1 + \mathrm{Tr}(\rho^2))) = (\rho + \rho^2)/(1 + \mathrm{Tr}(\rho^2))$.
For depolarized states $\rho(\delta) = (1-\delta)\psi + \delta I/d$, recursive application yields error $\delta_k \sim \delta^{2^k}$---doubly exponential convergence~\cite{Childs2025}.

We note the structural similarity to entanglement purification~\cite{Bennett1996}, which also uses bilateral operations and post-selection on two copies; however, coherence purification operates on a single system and the relevant resource is $\ell_1$-coherence rather than entanglement.

Under depolarizing noise ($p = 0.3$), the standard recursive swap test achieves rapid convergence (Table~\ref{tab:standard_depol}).

\begin{table}[tb]
\caption{Standard recursive swap test under depolarizing noise ($p = 0.3$). $F_\mathrm{cat}$ values shown.}
\label{tab:standard_depol}
\centering\footnotesize
\setlength{\tabcolsep}{3pt}
\begin{ruledtabular}
\begin{tabular}{lcccc}
$d$ & $n = 8$ & $n = 32$ & $n = 64$ & $n^*$ (distillation) \\
\colrule
4 & 0.949 & 0.986 & 0.993 & $7.3 \times 10^5$ \\
8 & 0.940 & 0.983 & 0.991 & $1.8 \times 10^7$ \\
16 & 0.935 & 0.982 & --- & $2.9 \times 10^8$ \\
64 & 0.932 & --- & --- & $7.3 \times 10^{10}$ \\
\end{tabular}
\end{ruledtabular}
\end{table}

Under energy-dependent dephasing ($\gamma = 2$), the standard swap test fails severely (Table~\ref{tab:standard_deph}).

\begin{table}[tb]
\caption{Standard recursive swap test under dephasing ($\gamma = 2$). $F_\mathrm{cat}$ at maximum tested copy count $n_\mathrm{max}$.}
\label{tab:standard_deph}
\centering\footnotesize
\setlength{\tabcolsep}{3pt}
\begin{ruledtabular}
\begin{tabular}{lccc}
Algorithm & $d$ & $F_\mathrm{cat}$ ($n_\mathrm{max}$) & $C_{\ell_1}$ \\
\colrule
QKAN & 4 & 0.438 (64) & 1.52 \\
qDRIFT & 8 & 0.195 (64) & 2.94 \\
CF-QPE & 16 & 0.088 (32) & 4.54 \\
Regev & 64 & 0.021 (8) & 12.5 \\
\end{tabular}
\end{ruledtabular}
\end{table}

At $d = 64$, $F_\mathrm{cat} = 0.021$ is marginally above $1/d \approx 0.016$ (random guessing).
The SWAP operator mixes states across different energy sectors, destroying the structure that dephasing preserves.
A \emph{covariant} swap test that replaces the full SWAP with sector-wise EC rotations $U_\mathrm{cov} = \bigoplus_{E} U_\mathrm{EC}^{(E)}(\pi/4)$ preserves energy-sector structure.
The covariant variant provides a modest improvement for $d \geq 8$ ($1.2$--$1.4\times$ in $F_\mathrm{cat}$), though for $d = 4$ the benefit is within numerical noise and the standard test can slightly outperform (Table~\ref{tab:unified_comparison}).
Neither variant alone is sufficient for practical recovery under strong dephasing.

\subsection{Dynamical decoupling and Clifford twirling}
\label{sec:dd_twirl}

The root cause of purification failure under dephasing is that dephasing is \emph{anisotropic}: it suppresses coherences $\rho_{ij}$ proportionally to $e^{-\gamma|E_i - E_j|}$, creating an exponentially non-uniform error profile that the swap test's symmetric projection cannot efficiently correct.
We introduce a three-stage pipeline that overcomes this limitation.

\textbf{CPMG dynamical decoupling.}
The CPMG sequence~\cite{MeiboomGill1958,Viola1999} applies $N$ equally spaced $\pi$-pulses during the dephasing interval.
We model its effect phenomenologically as
\begin{equation}
\gamma_\mathrm{eff} = \frac{\gamma}{N + 1}.
\label{eq:dd_gamma}
\end{equation}
Two caveats are essential.
First, Eq.~\eqref{eq:dd_gamma} is an \emph{assumed} suppression law, not a derived one: for strictly memoryless (Markovian) dephasing, $\pi$-pulse conjugation leaves the dephasing dissipator exactly invariant ($XZX = -Z$ implies $Z\rho Z$ is unchanged), so ideal DD provides \emph{no} suppression in that regime; genuine CPMG suppression requires temporally correlated (quasi-static or spectrally structured) noise, where filter-function theory gives model-dependent scalings~\cite{Viola1999}.
Second, our simulations contain no pulse-level dynamics: the DD stage is implemented as the parameter substitution $\gamma \to \gamma/(N+1)$ applied to a static dephasing map.
Equation~\eqref{eq:dd_gamma} should therefore be read as defining the noise-suppression budget the pipeline assumes is available from DD on a bath with suitable temporal correlations, and all downstream pipeline fidelities are conditional on that budget being achievable on hardware.
Table~\ref{tab:dd_sweep} shows the effect of the assumed CPMG pulse count on catalyst fidelity.

\begin{table}[tb]
\caption{DD sweep for qDRIFT ($d = 8$, $n = 8$ copies, $\gamma = 2$). $\gamma_\mathrm{eff}$ is the effective dephasing after DD, $p_\mathrm{eff}$ is the depolarizing parameter after twirling, and $F_\mathrm{cat}$ is the catalyst fidelity after the full pipeline.}
\label{tab:dd_sweep}
\centering\footnotesize
\setlength{\tabcolsep}{3pt}
\begin{ruledtabular}
\begin{tabular}{lccc}
DD config & $\gamma_\mathrm{eff}$ & $p_\mathrm{eff}$ & $F_\mathrm{cat}$ \\
\colrule
No DD & 2.000 & 0.961 & 0.176 \\
CPMG-2 & 0.667 & 0.541 & 0.800 \\
CPMG-4 & 0.400 & 0.366 & 0.914 \\
CPMG-8 & 0.222 & 0.221 & 0.963 \\
CPMG-16 & 0.118 & 0.123 & 0.983 \\
\end{tabular}
\end{ruledtabular}
\end{table}

\begin{figure}[tb]
\centering
\includegraphics[width=\columnwidth]{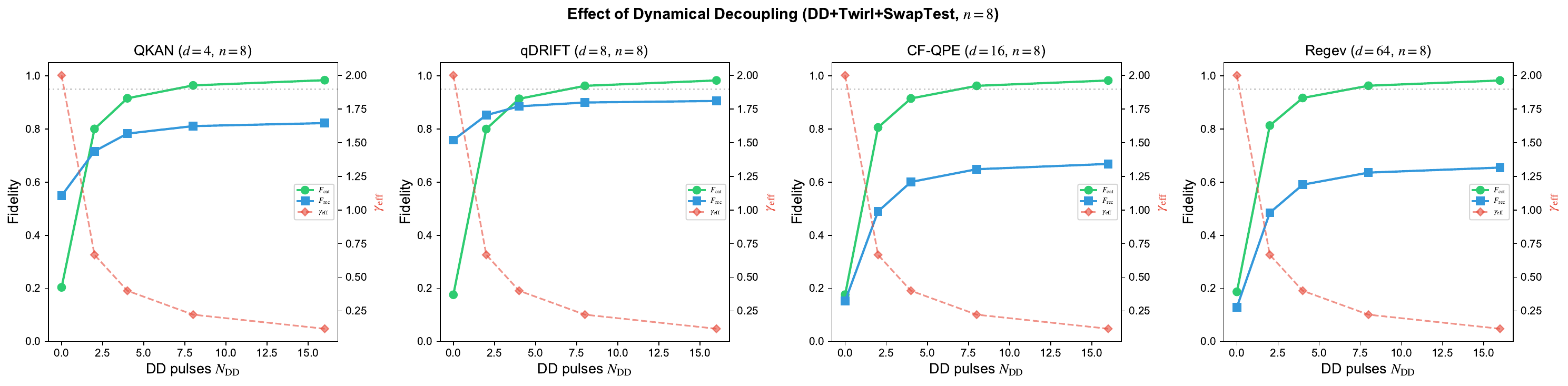}
\caption{Catalyst fidelity vs.\ CPMG pulse count for qDRIFT ($d = 8$, $n = 8$ copies, $\gamma = 2$). DD reduces effective dephasing, twirling isotropizes, and the swap test purifies efficiently.}
\label{fig:dd_sweep}
\end{figure}

Even moderate DD pulse counts produce dramatic improvements: CPMG-2 lifts $F_\mathrm{cat}$ from 0.176 to 0.800, and CPMG-8 achieves $F_\mathrm{cat} = 0.963$ (Fig.~\ref{fig:dd_sweep}).
The same trend holds across all tested dimensions: at CPMG-8, $F_\mathrm{cat} = 0.964$ ($d = 4$), $0.963$ ($d = 8$, $16$), and $0.963$ ($d = 64$), confirming dimension independence (Table~\ref{tab:dd_pipeline}).

\textbf{Clifford twirling.}
After DD reduces $\gamma$ to $\gamma_\mathrm{eff}$, the residual noise is still anisotropic dephasing.
Clifford twirling~\cite{Dankert2009,Wallman2016} converts any noise channel into a depolarizing channel by averaging over random Clifford gates (a technique related to randomized benchmarking~\cite{Dankert2009} and randomized compiling~\cite{Wallman2016}):
\begin{equation}
\overline{\mathcal{E}}_\mathrm{twirl}(\rho) = (1 - p_\mathrm{eff})\rho + p_\mathrm{eff}\, \frac{I}{d},
\label{eq:twirl}
\end{equation}
where
\begin{equation}
p_\mathrm{eff} = 1 - \frac{d\, \bar{F}_\mathrm{avg} - 1}{d - 1},
\label{eq:p_eff}
\end{equation}
and we compute $\bar{F}_\mathrm{avg} = e^{-\gamma_\mathrm{eff}} + (1 - e^{-\gamma_\mathrm{eff}})\,\frac{2}{d(d+1)}$, a uniform-attenuation approximation in which every coherence is damped by the slowest-mode factor $e^{-\gamma_\mathrm{eff}}$; its $d$-dependence vanishes as $d$ grows, giving $p_\mathrm{eff} \approx 1 - e^{-\gamma_\mathrm{eff}}$ (0.239/0.221/0.211/0.202 for $d = 4/8/16/64$ at $\gamma_\mathrm{eff} = 2/9$).
We flag an important sensitivity: if $\bar{F}_\mathrm{avg}$ is instead computed mode-resolved over the energy ladder, $\bar{F}_\mathrm{avg} = d^{-2}\sum_{i,j} e^{-\gamma_\mathrm{eff}|E_i - E_j|}$ with $H_S = \mathrm{diag}(0, \ldots, d-1)$, one obtains strongly dimension-dependent $p_\mathrm{eff} = 0.30/0.45/0.63/0.88$ for the same $d$ values, and the swap test would fail to purify at $d = 64$.
The dimension independence reported below is therefore a property of the uniform-attenuation twirl model, and identifying which model describes a given hardware channel is a prerequisite for the pipeline's large-$d$ claims.
In our simulations we use the exact analytical twirled channel rather than sampled Cliffords; a finite-sample implementation with $K$ random Cliffords would add a sampling error that our simulations do not capture.

\subsection{Combined DD+Twirl+Swap~Test pipeline}
\label{sec:pipeline}

The full pipeline operates in three stages:
\begin{enumerate}
\item \textbf{DD stage}: Apply CPMG-$N$ to each noisy copy, reducing $\gamma \to \gamma_\mathrm{eff} = \gamma/(N+1)$.
\item \textbf{Twirl stage}: Apply Clifford twirling to each DD-protected copy, converting dephasing($\gamma_\mathrm{eff}$) $\to$ depolarizing($p_\mathrm{eff}$).
\item \textbf{Purification stage}: Apply the standard recursive swap test, exploiting doubly exponential convergence under depolarizing noise.
\end{enumerate}

The physical interpretation is: DD suppresses the noise magnitude, twirling isotropizes the residual noise geometry, and the swap test purifies the isotropic noise efficiently.
Each stage addresses a distinct obstacle---magnitude, anisotropy, and mixedness---and the composition is more effective than any single technique.

Table~\ref{tab:dd_pipeline} presents the main results with CPMG-8 and $n = 8$ copies.

\begin{table}[tb]
\caption{DD+Twirl+Swap~Test pipeline results (CPMG-8, $\gamma_\mathrm{eff} = 0.222$, $n = 8$ copies, original $\gamma = 2$). ``Raw $F_\mathrm{cat}$'' is without DD or twirling. Catalyst-preparation rows use the maximally coherent reference state of the corresponding dimension $d$ (the catalyst target); the algorithm names label the dimension and the downstream $F_\mathrm{rec}$ benchmark, not the purified state.}
\label{tab:dd_pipeline}
\centering\footnotesize
\setlength{\tabcolsep}{3pt}
\begin{ruledtabular}
\begin{tabular}{lcccc}
Algorithm & $d$ & Raw $F_\mathrm{cat}$ & \makecell{DD+Twirl \\ $F_\mathrm{cat}$} & \makecell{DD+Twirl \\ $F_\mathrm{rec}$} \\
\colrule
QKAN & 4 & 0.351 & 0.964 & 0.811 \\
qDRIFT & 8 & 0.173 & 0.963 & 0.900 \\
CF-QPE & 16 & 0.085 & 0.963 & 0.649 \\
Regev & 64 & 0.021 & 0.963 & 0.636 \\
\end{tabular}
\end{ruledtabular}
\end{table}

\begin{figure}[tb]
\centering
\includegraphics[width=\columnwidth]{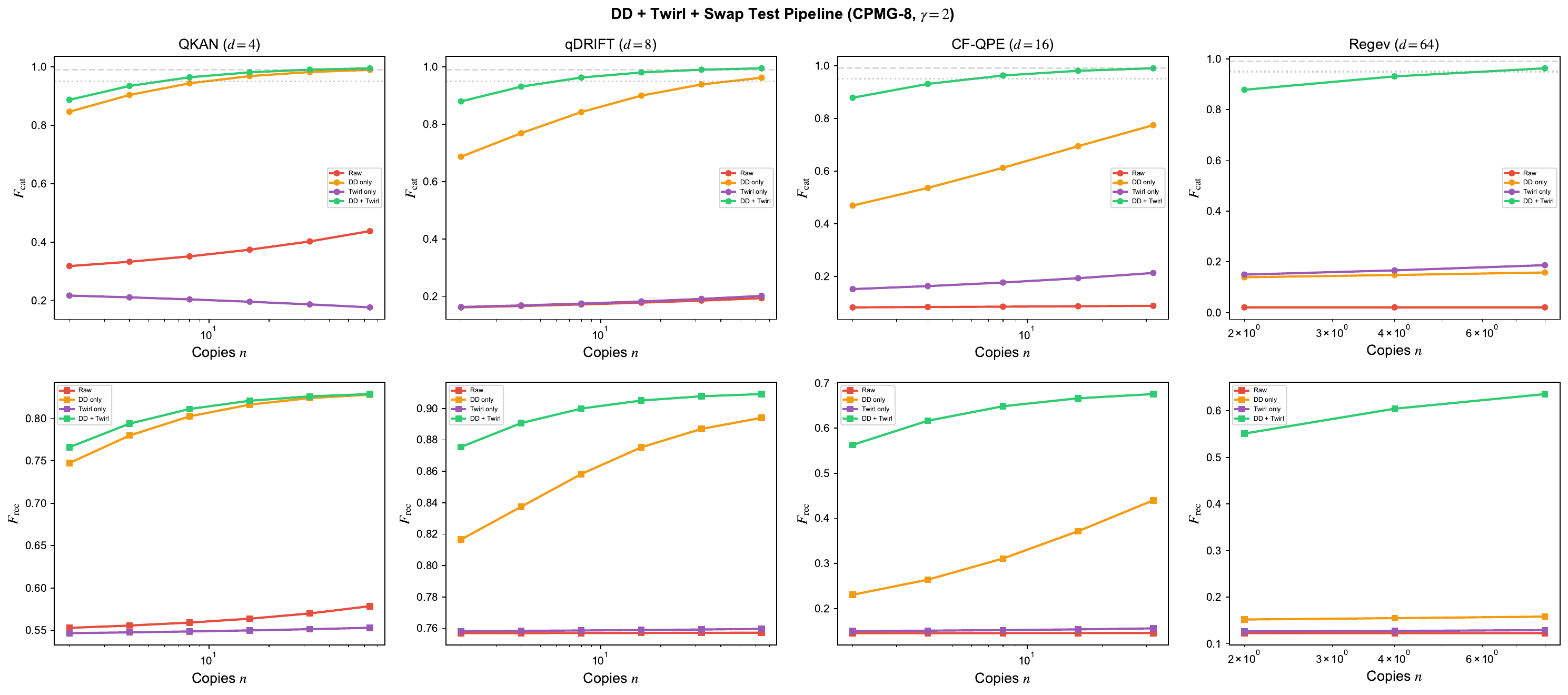}
\caption{\textbf{The DD+Twirl+Swap-Test pipeline achieves $F_\mathrm{cat} > 0.96$ uniformly across $d = 4$--$64$ from only 8 noisy copies, whereas no single stage suffices.}
Comparison under dephasing $\gamma = 2$ with $n = 8$ copies and CPMG-8.
Raw purification (red) fails for $d \geq 8$; DD alone (orange) improves but remains dimension-dependent; twirl alone (purple) fails; combined DD+Twirl (green) achieves dimension-independent recovery.}
\label{fig:dd_pipeline}
\end{figure}

\begin{figure}[tb]
\centering
\includegraphics[width=\columnwidth]{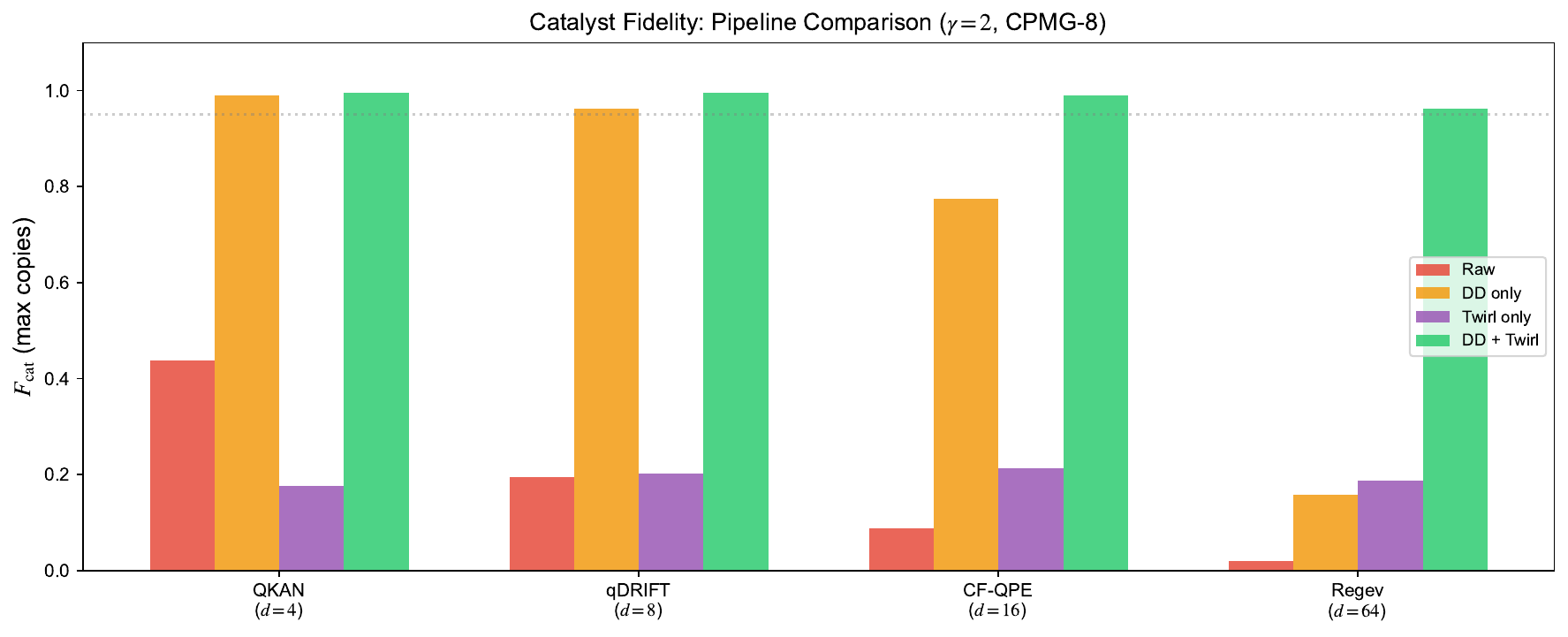}
\caption{Summary of all catalyst preparation strategies under dephasing $\gamma = 2$: variational (0 copies), standard swap test, covariant swap test, and DD+Twirl+Swap~Test pipeline (CPMG-8, 8 copies). The DD+Twirl pipeline dramatically outperforms all other methods.}
\label{fig:dd_summary}
\end{figure}

The pipeline achieves $F_\mathrm{cat} > 0.96$ uniformly across all dimensions, in stark contrast to the raw swap test ($F_\mathrm{cat} = 0.02$--$0.44$), DD alone (dimension-dependent), or twirling alone (which fails because twirling without DD converts strong dephasing into strong depolarizing noise).
An ablation analysis confirms that each stage is necessary:
DD alone reduces $\gamma$ but leaves anisotropic noise ($F_\mathrm{cat} \approx 0.6$--$0.8$, dimension-dependent);
twirling alone converts strong dephasing into strong depolarizing ($p_\mathrm{eff} \approx 0.96$ for $\gamma = 2$), which the swap test cannot efficiently purify;
DD+Twirl without the swap test produces $F_\mathrm{cat} \approx 0.85$ (the twirled state's overlap with the target);
and the full pipeline achieves $F_\mathrm{cat} > 0.96$ uniformly (Figs.~\ref{fig:dd_pipeline} and~\ref{fig:dd_summary}).

The per-copy resource cost is $N$ DD pulses (CPMG-8: 8 $\pi$-pulses) and $O(n_q^2)$ Clifford gates for twirling, with $n = 2^k$ input copies for the swap test.
For a like-for-like comparison we match the fidelity endpoint of the distillation baseline ($F_\mathrm{cat} \geq 0.99$), which the pipeline reaches at $n = 32$ copies: the copy-count reduction is then $7.3 \times 10^{5}/32 \approx 2 \times 10^{4}$ ($d = 4$), $1.8 \times 10^{7}/32 \approx 6 \times 10^{5}$ ($d = 8$), $2.9 \times 10^{8}/32 \approx 9 \times 10^{6}$ ($d = 16$), and $7.3 \times 10^{10}/32 \approx 2 \times 10^{9}$ ($d = 64$, the high-dimensional stress case)---i.e., four to nine orders of magnitude depending on dimension.
At the looser endpoint $F_\mathrm{cat} = 0.963$, eight copies suffice.
Two caveats on this accounting: (i)~it counts conditional input copies---the swap-test cascade is postselected, and including the measured success probability ($P_\mathrm{succ} \approx 0.68$--$0.71$ for three rounds across $d = 4$--$64$ in our simulations) raises the expected copy consumption to $\approx 11$--$12$ per successful catalyst at $n = 8$; and (ii)~both the pipeline and the distillation baseline require one algorithm-preparation circuit per copy, so per-copy gate overheads largely cancel and the copy ratio is the meaningful comparison.

\section{Benchmark Algorithms and Decoherence Models}
\label{sec:benchmarks}

\subsection{Quantum algorithms}
\label{sec:algorithms}

We test CQEC on four algorithms chosen to span the major application domains and cover a range of Hilbert space dimensions ($d = 4$--$64$), coherence structures, and algorithmic noise profiles.

\textbf{1.\ qDRIFT}~\cite{Campbell2019,Chen2021} simulates Hamiltonian dynamics $e^{-iHt}$ for the 3-qubit Heisenberg model $H = J \sum_{\langle i,j \rangle} (\sigma_i^x \sigma_j^x + \sigma_i^y \sigma_j^y + \sigma_i^z \sigma_j^z) + h \sum_i \sigma_i^z$ with $J = 1.0$, $h = 0.5$, $t = 1.0$.
The qDRIFT approximation uses 80 random product formula gates with probabilistic sampling~\cite{Campbell2019}.
Dimension $d = 8$ (3~qubits).
Since different random seeds produce different gate sequences, $F_\mathrm{before}$ varies across seeds (std $\approx 0.17$ for dephasing $\gamma = 2$, reflecting the randomized product formula's state-dependent approximation error).
Crucially, the \emph{post-correction} fidelity $F_\mathrm{after}$ shows negligible seed dependence (std $< 10^{-4}$), confirming that CQEC is robust to the input state's seed-dependent structure.

\textbf{2.\ QKAN}~\cite{Ivashkov2024} implements a quantum Kolmogorov--Arnold network layer encoding the first four Chebyshev polynomials $T_n(x)$ at $x = 0.5$: the ideal state has amplitudes proportional to $(T_0, T_1, T_2, T_3) = (1.0, 0.5, -0.5, -1.0)$ (normalized), while the algorithmic output truncates at degree~2.
The CQEC target is the algorithmic output (not the ideal state), so recovery corrects only decoherence.
Dimension $d = 4$ (2~qubits).

\textbf{3.\ Control-free QPE}~\cite{Clinton2026} estimates eigenvalues of a Fermi--Hubbard-type Hamiltonian using vectorial phase retrieval without controlled unitaries.
The protocol generates time series $f_j = \braket{\psi | e^{-iHj\Delta t} | \psi}$ and encodes the resulting spectrum as a 16-dimensional quantum state.
Dimension $d = 16$ (4~qubits).

\textbf{4.\ Regev factoring}~\cite{Regev2024} factors $N = 15$ using discrete Gaussian states with $D = 8$ grid points per dimension, modular exponentiation, and quantum Fourier transform for LLL reduction~\cite{LLL1982}.
Dimension $d = 64$ (6~qubits).

\subsection{Decoherence models}
\label{sec:noise}

Three noise channels are applied to each algorithm's output state:

\textbf{Partial dephasing.}
$\mathcal{E}_\mathrm{deph}(\rho)_{ij} = \rho_{ij} \cdot e^{-\gamma|E_i - E_j|}$, with $\gamma = 2.0$ (strong dephasing).
Two conventions labeled ``dephasing $\gamma$'' appear in this paper: the energy-ladder form above (used in the asymptotic benchmarks) and a per-qubit $Z$-channel with uniform attenuation $e^{-\gamma}$ on all coherences (used in the finite-$n$, entanglement, and pipeline sections); pre-correction fidelities therefore differ between sections for nominally equal $\gamma$, and each table states which convention applies.
This preserves all coherent modes for $\gamma < \infty$: $\mathcal{D}(\rho_\mathrm{noisy}) = \mathcal{D}(\rho)$.
For dephased states, the pre-correction fidelity with the ideal state is
\begin{align}
F_\mathrm{before}(\gamma) &= \mathrm{Tr}[\rho_\mathrm{diag}^2] + e^{-\gamma}\bigl(1 - \mathrm{Tr}[\rho_\mathrm{diag}^2]\bigr),
\label{eq:fidelity_decay}
\end{align}
which approaches $1/d$ as $\gamma \to \infty$ for the maximally coherent state.

\textbf{Depolarizing.}
$\mathcal{E}_\mathrm{depol}(\rho) = (1-p)\rho + p \cdot I/d$, with $p = 0.3$.
Since $I/d$ has zero off-diagonal elements, all coherent modes are preserved for $p < 1$.

\textbf{Combined.}
Sequential: dephasing ($\gamma = 1.0$), depolarizing ($p = 0.15$), amplitude damping ($\gamma_\mathrm{AD} = 0.1$).
The amplitude damping channel with Kraus operators $E_0 = \ket{0}\!\bra{0} + \sqrt{1-\gamma_\mathrm{AD}}\,\ket{1}\!\bra{1}$, $E_1 = \sqrt{\gamma_\mathrm{AD}}\,\ket{0}\!\bra{1}$ is \emph{not} a covariant channel but preserves all coherent modes for $\gamma_\mathrm{AD} < 1$.
The CQEC \emph{recovery} channel $\Lambda$ is covariant, while the \emph{noise} channel need not be; the key requirement is that the noise preserves the mode support $\mathcal{D}(\rho)$.

\subsection{TTN cryptographic protocol}
\label{sec:ttn}

To demonstrate CQEC's applicability beyond algorithm protection, we test it on quantum states generated by a tree tensor network (TTN) cryptographic protocol~\cite{Huggins2019,Sim2019}.
We simulate a 3-qubit reduced model with layered RY--CNOT--RZ structure, tested across 6 plaintext lengths ($N_c = 5, 10, 15, 20, 25, 30$) and 7 noise levels from ideal to extreme.

\subsection{Conventional QEC baselines}
\label{sec:qec_baselines}

For comparison, we compute the logical fidelity of:
\begin{itemize}
\item \textbf{Steane $[\![7,1,3]\!]$ code}~\cite{Steane1996}: corrects any single-qubit error ($t = 1$). $F_L = \sum_{k=0}^{t} \binom{7}{k} p_\mathrm{eff}^k (1-p_\mathrm{eff})^{7-k}$.
\item \textbf{Surface code}~\cite{Kitaev2003,Fowler2012} at distances $d = 3$ and $d = 5$: below threshold we use the standard scaling $p_L \sim (p/p_\mathrm{th})^{(d+1)/2}$ with $p_\mathrm{th} \approx 0.01$; above threshold, where this asymptotic expression no longer applies, the tabulated values use a saturating binomial model over the $d^2$ physical qubits (logical failure when more than $\lfloor d/2 \rfloor$ physical errors occur), clamped to $[0, 1]$.
\end{itemize}
The effective per-qubit error rate is $p_\mathrm{eff} = 1 - (1-p)^{1/n_q}$.
These are analytical models; both sides of the comparison represent idealized performance.
The comparison illustrates qualitative differences: threshold-dependent degradation vs.\ threshold-free recovery.
We emphasize that QEC codes correct \emph{arbitrary} errors up to weight $t$ on \emph{unknown} encoded states, while CQEC recovers \emph{known} states from arbitrary noise strength---these are complementary capabilities, not competing ones.

\section{Results}
\label{sec:results}

\subsection{CQEC asymptotic benchmark}
\label{sec:main_results}

We present results in two regimes: the \emph{asymptotic limit} ($n \to \infty$, Secs.~\ref{sec:main_results}--\ref{sec:durability}), which validates the existence and robustness of catalytic recovery, and the \emph{finite-copy regime} (Secs.~\ref{sec:finite_copy}--\ref{sec:scaling}), which quantifies the practical resource cost.
We caution that the asymptotic results represent an upper bound on achievable performance; Table~\ref{tab:finite_n} gives the copy counts needed to approach these fidelities, and the DD+Twirl pipeline (Sec.~\ref{sec:dd_benchmarks}) dramatically reduces these costs.
The two regimes should be read together: Sec.~\ref{sec:main_results} answers ``what is possible?'' and Secs.~\ref{sec:finite_copy}--\ref{sec:scaling} answer ``at what cost?''

Table~\ref{tab:main} presents CQEC recovery results in the asymptotic limit from 10 independent trials.
Regev ($d = 64$) achieves slightly lower $F_\mathrm{after}$ compared to smaller systems, reflecting the reduced expressibility of the 5-parameter ansatz in the larger Hilbert space.

\begin{table*}[t]
\caption{Asymptotic recovery fidelity in the \emph{effective model} (mean $\pm$ std, 10~seeds), generated by the legacy effective-map implementation (see Sec.~\ref{sec:optimization} and Code availability); these values are not reproduced by the packaged finite-$n$ variational path, which yields $F \approx 0.86$--$0.94$. Success rate is 100\% for all entries. Regev results (below rule) serve as a high-dimensional stress test; CQEC is inapplicable to factoring in practice due to the target-state requirement (Sec.~\ref{sec:limitations}).}
\label{tab:main}
\begin{ruledtabular}
\begin{tabular}{llcc}
Algorithm & Noise model & $F_\mathrm{before}$ & $F_\mathrm{after}$ \\
\colrule
qDRIFT (3~qb) & Dephasing ($\gamma=2$) & $0.701 \pm 0.171$ & $0.9999 \pm 0.0000$ \\
qDRIFT (3~qb) & Depolarizing ($p=0.3$) & $0.528 \pm 0.119$ & $0.9997 \pm 0.0001$ \\
qDRIFT (3~qb) & Combined & $0.615 \pm 0.145$ & $0.9999 \pm 0.0001$ \\
QKAN (2~qb) & Dephasing ($\gamma=2$) & $0.341 \pm 0.000$ & $1.0000 \pm 0.0000$ \\
QKAN (2~qb) & Depolarizing ($p=0.3$) & $0.495 \pm 0.000$ & $1.0000 \pm 0.0000$ \\
QKAN (2~qb) & Combined & $0.386 \pm 0.000$ & $1.0000 \pm 0.0000$ \\
CF-QPE (4~qb) & Dephasing ($\gamma=2$) & $0.306 \pm 0.000$ & $1.0000 \pm 0.0000$ \\
CF-QPE (4~qb) & Depolarizing ($p=0.3$) & $0.718 \pm 0.000$ & $1.0000 \pm 0.0000$ \\
CF-QPE (4~qb) & Combined & $0.427 \pm 0.000$ & $1.0000 \pm 0.0000$ \\
TTN-Crypto (3~qb) & Dephasing ($\gamma=2$) & $0.388 \pm 0.067$ & $1.0000 \pm 0.0000$ \\
TTN-Crypto (3~qb) & Depolarizing ($p=0.3$) & $0.734 \pm 0.002$ & $1.0000 \pm 0.0000$ \\
TTN-Crypto (3~qb) & Combined & $0.487 \pm 0.042$ & $1.0000 \pm 0.0000$ \\
\colrule
Regev (6~qb) & Dephasing ($\gamma=2$) & $0.066 \pm 0.000$ & $0.9998 \pm 0.0000$ \\
Regev (6~qb) & Depolarizing ($p=0.3$) & $0.505 \pm 0.000$ & $0.9996 \pm 0.0001$ \\
Regev (6~qb) & Combined & $0.271 \pm 0.000$ & $0.9997 \pm 0.0001$ \\
\end{tabular}
\end{ruledtabular}
\end{table*}

\subsection{Sharp threshold}
\label{sec:threshold}

\begin{figure}[!tbp]
\centering
\includegraphics[width=\columnwidth]{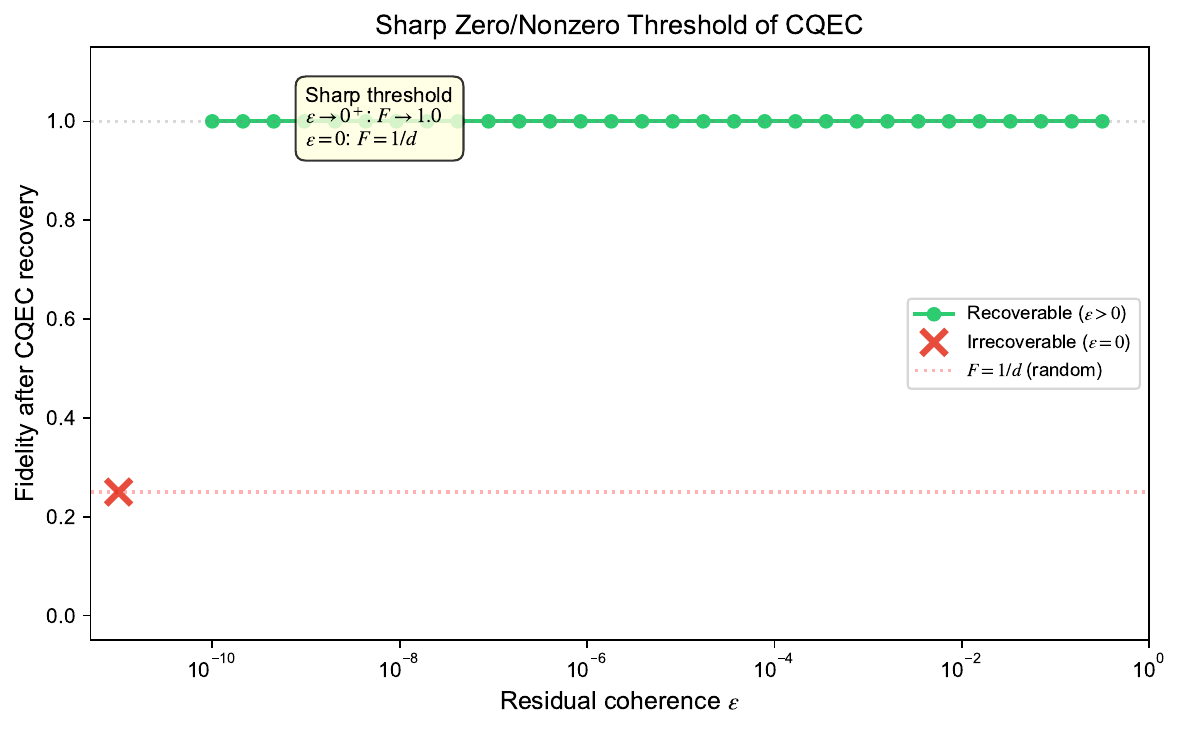}
\caption{\textbf{CQEC exhibits an infinitely sharp recovery threshold over 10~orders of magnitude in residual coherence.}
Post-correction fidelity is plotted against residual coherence $\varepsilon$ (log scale).
Green circles: successful recovery ($\varepsilon > 0$, $F \to 1$).
Red cross: failed recovery at $\varepsilon = 0$ ($F = 1/d$).
Dashed line: $F = 1/d = 0.25$ (random guess).}
\label{fig:threshold}
\end{figure}

Figure~\ref{fig:threshold} probes the zero/nonzero threshold predicted by Theorem~1, with two scope qualifications.
First, what is verified directly is the \emph{mode-inclusion classification}: for a 2-qubit ($d = 4$) maximally coherent target, mode inclusion holds for every tested $\varepsilon > 0$ (30~values on $[10^{-10}, 0.32]$) and fails at $\varepsilon = 0$; the accompanying recovery fidelities ($F_\mathrm{after} = 1.000$ for $\varepsilon > 0$, $1/d$ at $\varepsilon = 0$) are outputs of the effective recovery model (Sec.~\ref{sec:optimization}) in the asymptotic setting.
Second, the numerical discontinuity necessarily sits at the mode-detection tolerance ($\epsilon_\mathrm{mode} = 10^{-14}$, with a $10^{-15}$ guard in the packaged implementation) rather than at the mathematical point $\varepsilon = 0$: coherences below the tolerance are treated as absent, so the scan range $[10^{-10}, 0.32]$ lies entirely within the resolved regime, and the ``infinitely sharp'' language refers to the theorem, not to a numerically resolved discontinuity.
The QFI tracks the classification: at $\varepsilon = 10^{-10}$, $\mathcal{F}(\rho_\mathrm{noisy}, H) = 3.2 \times 10^{-19}$ yet the state is classified recoverable, while at $\varepsilon = 0$, $\mathcal{F} = 0$ and it is not.
Within the model, the threshold is controlled by the \emph{support} of the coherence, not its magnitude.

We also tested a \emph{selective dephasing} channel that destroys the $\Delta_{01}$ mode while preserving all others: CQEC achieves only $F_\mathrm{after} = 0.72$ (vs.\ $F_\mathrm{after} = 1.0$ with all modes present).
The intermediate value $F = 0.72 > 1/d = 0.25$ reflects partial recovery: the optimizer finds a covariant map that reconstructs the state using the surviving modes, but cannot restore the lost $\Delta_{01}$ component.
This demonstrates that mode inclusion violations produce graceful degradation rather than catastrophic failure.

\subsection{Noise strength sweep}
\label{sec:noise_sweep}

\begin{figure}[!tbp]
\centering
\includegraphics[width=\columnwidth]{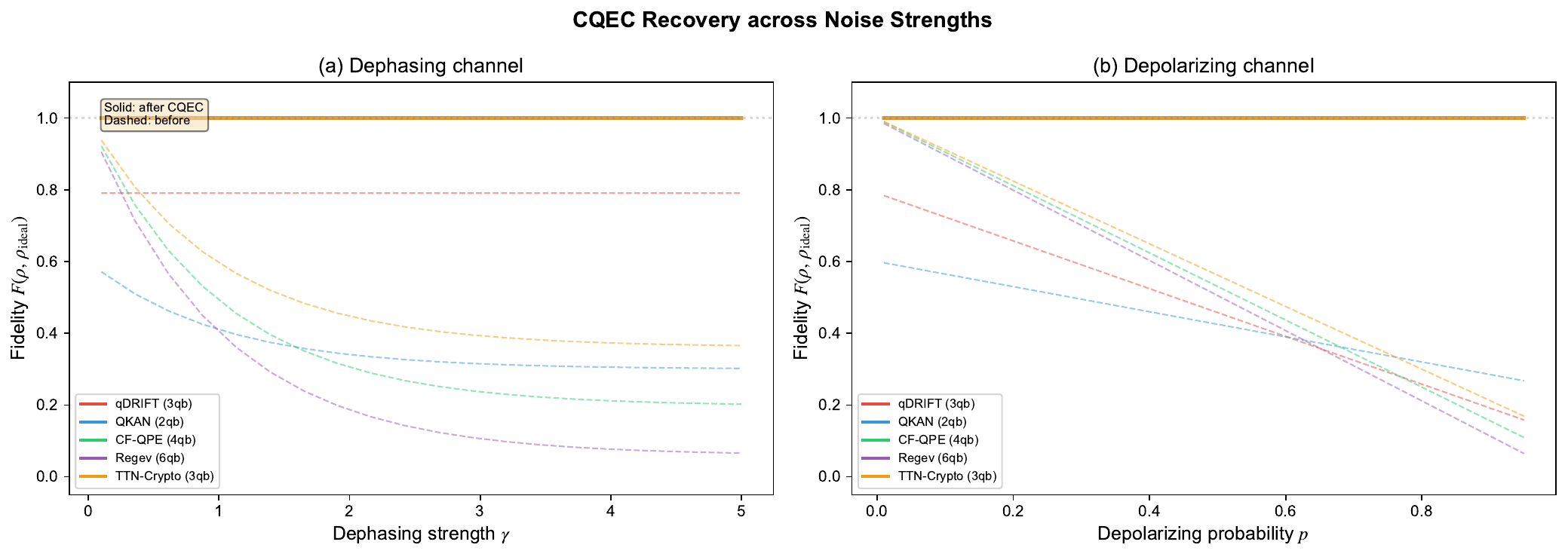}
\caption{\textbf{CQEC maintains $F \approx 1$ across the full range of dephasing and depolarizing noise.}
(a)~Dephasing sweep over $\gamma \in [0.1, 5.0]$ (20~points).
(b)~Depolarizing sweep over $p \in [0.01, 0.95]$ (20~points).
Dashed lines: pre-correction fidelity; solid lines: post-CQEC fidelity.}
\label{fig:noise_sweep}
\end{figure}

Figure~\ref{fig:noise_sweep} presents CQEC recovery fidelity vs.\ noise strength.
For dephasing with $\gamma \in [0.1, 5.0]$ (20~points) and depolarizing with $p \in [0.01, 0.95]$ (20~points), CQEC maintains $F_\mathrm{after} > 0.99$ across all $5 \times 2 \times 20 = 200$ data points, even when $F_\mathrm{before}$ drops to 0.066 (Regev, $\gamma = 5.0$) or 0.065 (Regev, $p = 0.95$).

The observed pre-correction decay is consistent with Eq.~\eqref{eq:fidelity_decay}: Regev ($d = 64$, $1/d = 0.016$) decays most steeply, while qDRIFT ($d = 8$, $1/d = 0.125$) decays more slowly due to non-uniform population distribution.
Post-CQEC fidelity remains flat at $F \approx 1.0$ regardless of $d$ or noise strength.
At $p = 1$ (complete depolarizing), $\rho_\mathrm{noisy} = I/d$ and all off-diagonal elements vanish, so $\mathcal{D}(\rho_\mathrm{noisy}) = \emptyset$ and recovery is impossible; the sweep stops at $p = 0.95$ where modes are still preserved ($|\rho_{ij}| \geq 0.05 \cdot |\rho_{ij}^{(0)}|$).

\subsection{Comparison with conventional QEC}
\label{sec:qec_comparison}

\begin{figure}[!tbp]
\centering
\includegraphics[width=\columnwidth]{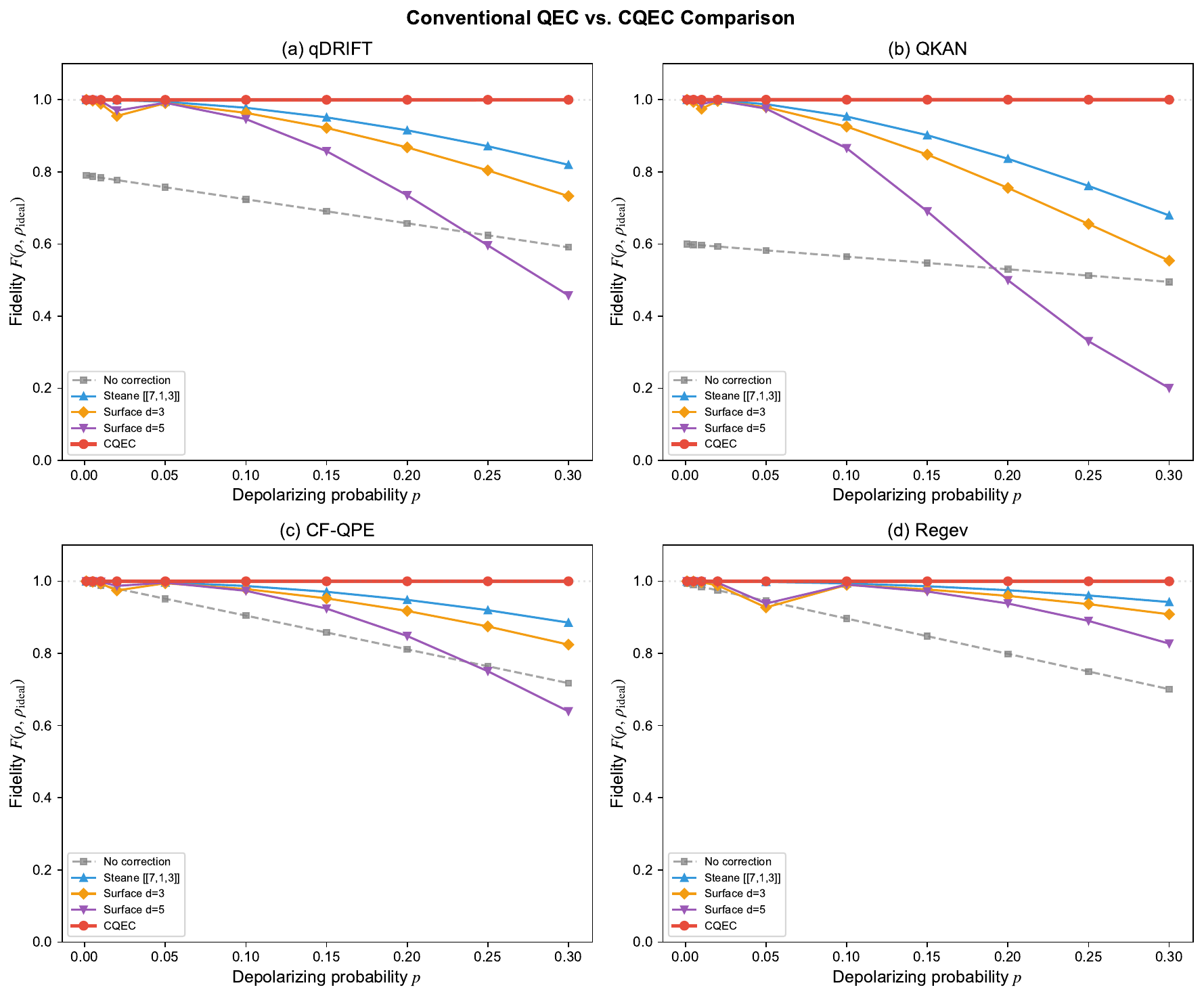}
\caption{\textbf{CQEC maintains unit fidelity at noise levels where Steane and surface codes degrade monotonically.}
Panels (a)--(d): qDRIFT, QKAN, CF-QPE, and Regev under depolarizing noise.
At $p = 0.3$, surface code $d=5$ falls to $F \approx 0.64$ while CQEC remains at $F = 1.0$.}
\label{fig:qec_comparison}
\end{figure}

Figure~\ref{fig:qec_comparison} compares CQEC against three conventional codes across depolarizing noise $p \in [0.001, 0.3]$.

\begin{table}[!tbp]
\caption{Fidelity comparison at representative error rates (CF-QPE, 4~qubits). CQEC values are asymptotic ($n \to \infty$); QEC values assume a single correction round with perfect syndrome extraction. Both represent idealized performance under their respective operational assumptions.}
\label{tab:qec}
\centering\footnotesize
\setlength{\tabcolsep}{3pt}
\begin{ruledtabular}
\begin{tabular}{lcccc@{\hspace{6pt}}c}
$p$ & None & Steane & Surf.\ $d\!=\!3$ & Surf.\ $d\!=\!5$ & CQEC \\
\colrule
0.01 & 0.989 & 1.000 & 0.994 & 0.998 & 1.000 \\
0.10 & 0.905 & 0.987 & 0.979 & 0.974 & 1.000 \\
0.20 & 0.811 & 0.949 & 0.918 & 0.848 & 1.000 \\
0.30 & 0.718 & 0.885 & 0.824 & 0.639 & 1.000 \\
\end{tabular}
\end{ruledtabular}
\end{table}

Key observations:
(1)~At $p = 0.01$ (below threshold), all codes perform well.
(2)~At $p = 0.1$ (above threshold for surface codes), surface code $d = 5$ degrades to 0.974 while CQEC remains at 1.000.
(3)~At $p = 0.3$, Steane falls to 0.885, surface $d = 5$ to 0.639, while CQEC maintains 1.000.
The pattern is consistent across all four algorithms (Fig.~\ref{fig:qec_comparison}a--d).

\subsection{TTN cryptographic protocol protection}
\label{sec:ttn_results}

\begin{figure}[!tbp]
\centering
\includegraphics[width=\columnwidth]{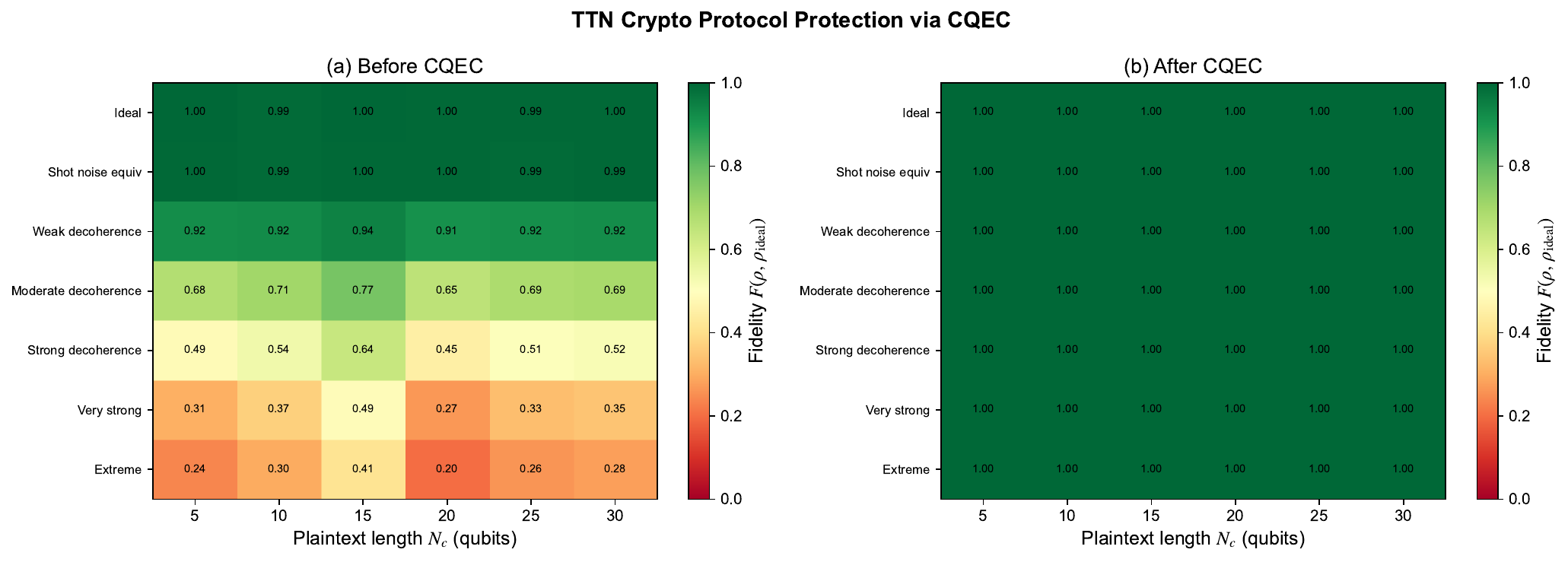}
\caption{TTN crypto protocol $+$ CQEC protection.
(a)~Fidelity before CQEC (heatmap by plaintext length and noise level).
(b)~Fidelity after CQEC (uniformly $F = 1.0$).}
\label{fig:ttn}
\end{figure}

Figure~\ref{fig:ttn} presents a heatmap across 42~data points (6 plaintext lengths $\times$ 7 noise levels).
We observe three regimes:
(i)~Weak noise ($\gamma \leq 0.5$, $p \leq 0.05$): pre-correction $F > 0.9$, CQEC provides marginal benefit.
(ii)~Moderate noise ($\gamma \sim 1.5$, $p \sim 0.15$): pre-correction $F \approx 0.5$--$0.7$, CQEC restores to $F = 1.000$.
(iii)~Extreme noise ($\gamma = 3.0$, $p = 0.3$): pre-correction $F = 0.20$--$0.28$, approaching $1/d = 0.125$; CQEC still achieves $F = 1.000$ in the asymptotic limit.
The recovery is uniform across all plaintext lengths, confirming that the mode structure of TTN output states is robust: all 6 plaintext lengths produce states with identical mode support $|\mathcal{D}| = 56$.
CQEC can restore intermediate quantum states in prepare-and-measure cryptographic schemes where the sender knows the target state, complementing the classical noise absorption mechanism in TTN protocols~\cite{Sim2019}.
The practical relevance is quantum key distribution with noisy channels: CQEC applied at the receiver side can restore quantum states before measurement, improving the raw key rate under high-loss conditions.

\subsection{Catalyst durability}
\label{sec:durability}

\begin{figure}[!tbp]
\centering
\includegraphics[width=\columnwidth]{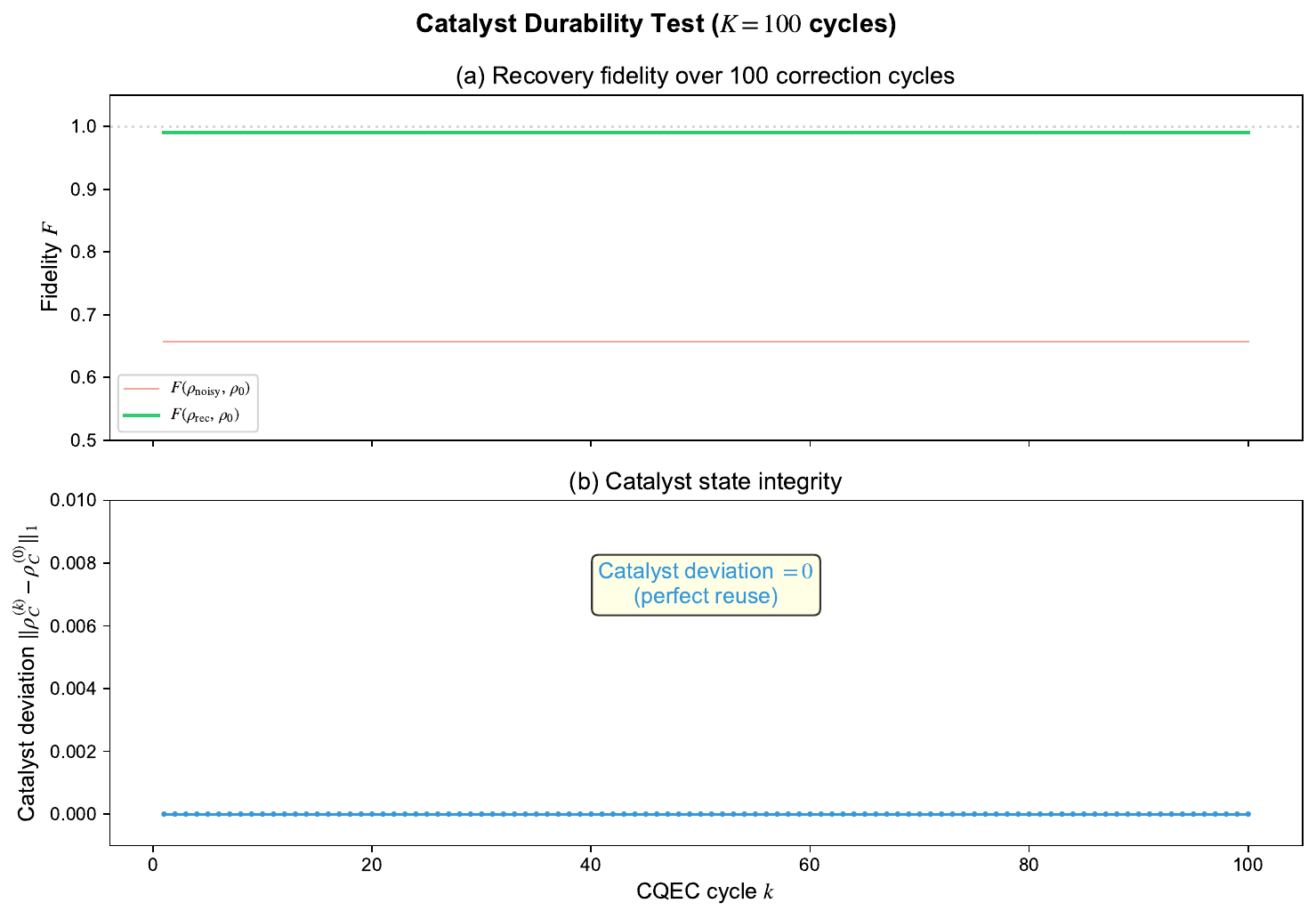}
\caption{\textbf{One-hundred-cycle consistency check of the effective implementation: recovery fidelity and catalyst state are numerically stable under deterministic replay.}
(a)~Recovery fidelity over 100~cycles (green: recovered, red: noisy).
(b)~Catalyst-state deviation from input ($< 10^{-12}$ per cycle).
Because the implementation carries the catalyst forward by construction, this figure verifies numerical stability, not Theorem~2's marginal-preservation property (see text).}
\label{fig:durability}
\end{figure}

Figure~\ref{fig:durability} shows 100 consecutive recovery cycles (qDRIFT, dephasing $\gamma = 1.5$, depolarizing $p = 0.2$), with the catalyst not re-initialized between cycles.
We report this as an \emph{internal consistency check}, not as evidence for Theorem~2's catalyst preservation: in the effective implementation the catalyst is carried forward by construction (its state is not obtained as the marginal of a propagated joint system--catalyst output), the same optimized parameters $\boldsymbol{\theta}$ are used in every cycle, and the noise realization is deterministic, so cycle-to-cycle stability ($F = 0.990$ in all 100~cycles; $\delta_c < 10^{-12}$ per cycle, $< 10^{-28}$ at 128-bit precision) is the expected behavior of a deterministic replay and the residual is floating-point roundoff.
A test with evidential force for correlated catalysis would require propagating the joint state, extracting the catalyst by partial trace, and resampling the noise each cycle; we leave this---and the accumulation of system--catalyst correlations it would expose---to the joint-simulation follow-up identified in Sec.~\ref{sec:optimization}.
We caution additionally that unlimited reuse is guaranteed only in the asymptotic limit; for finite-$n$ implementations, drift of order $O(1/\sqrt{n})$ per cycle could accumulate.

\subsection{Finite-copy fidelity bounds}
\label{sec:finite_copy}

Table~\ref{tab:finite_n} reports the copy number $n^*$ required to achieve target fidelities, obtained by solving the \emph{full} bound of Eq.~\eqref{eq:finite_n_fidelity}, $1 - F = C^2/(4n) + C/\sqrt{n}$, for $n$ (both terms retained; the $C/\sqrt{n}$ term dominates at these targets).
The constants $C$ are \emph{effective empirical values}: the rigorous bound $C \leq d\,C_{\ell_1}/\min|\rho_{ij}|$ [Eq.~\eqref{eq:C_bound}] evaluated on the quoted spectra gives much larger values ($\sim\!47$--$2.5 \times 10^6$), so the tabulated $C$ should be read as fitted effective constants rather than derived bounds, and the scaling fit below shares this caveat.

\begin{table}[!tbp]
\caption{Input copies $n^*$ for target fidelity under dephasing ($\gamma = 2.0$), from the full bound $1-F = C^2/(4n) + C/\sqrt{n}$. Values are consistent with Table~\ref{tab:bottleneck}.}
\label{tab:finite_n}
\centering\footnotesize
\setlength{\tabcolsep}{3pt}
\begin{ruledtabular}
\begin{tabular}{lccc}
Algorithm & $C$ & $n^*(F\!\geq\!0.99)$ & $n^*(F\!\geq\!0.999)$ \\
\colrule
qDRIFT ($d=8$) & 42 & $1.8 \times 10^7$ & $1.8 \times 10^9$ \\
QKAN ($d=4$) & 8.5 & $7.3 \times 10^5$ & $7.2 \times 10^7$ \\
CF-QPE ($d=16$) & 170 & $2.9 \times 10^8$ & $2.9 \times 10^{10}$ \\
Regev ($d=64$) & 2700 & $7.3 \times 10^{10}$ & $7.3 \times 10^{12}$ \\
\end{tabular}
\end{ruledtabular}
\end{table}

A log-log fit of the effective constants yields $C \approx 0.53\,d^{2.06}$, so $n^* \sim d^{4.1} e^{2\gamma}$ at fixed target fidelity; because the $C$ values are fitted rather than independently derived, this exponent is descriptive, not a validated scaling law.
Readers should interpret Table~\ref{tab:main} as ``what the effective model predicts asymptotically'' and Table~\ref{tab:finite_n} as ``what the constructive route costs.''

\begin{figure}[!tbp]
\centering
\includegraphics[width=\columnwidth]{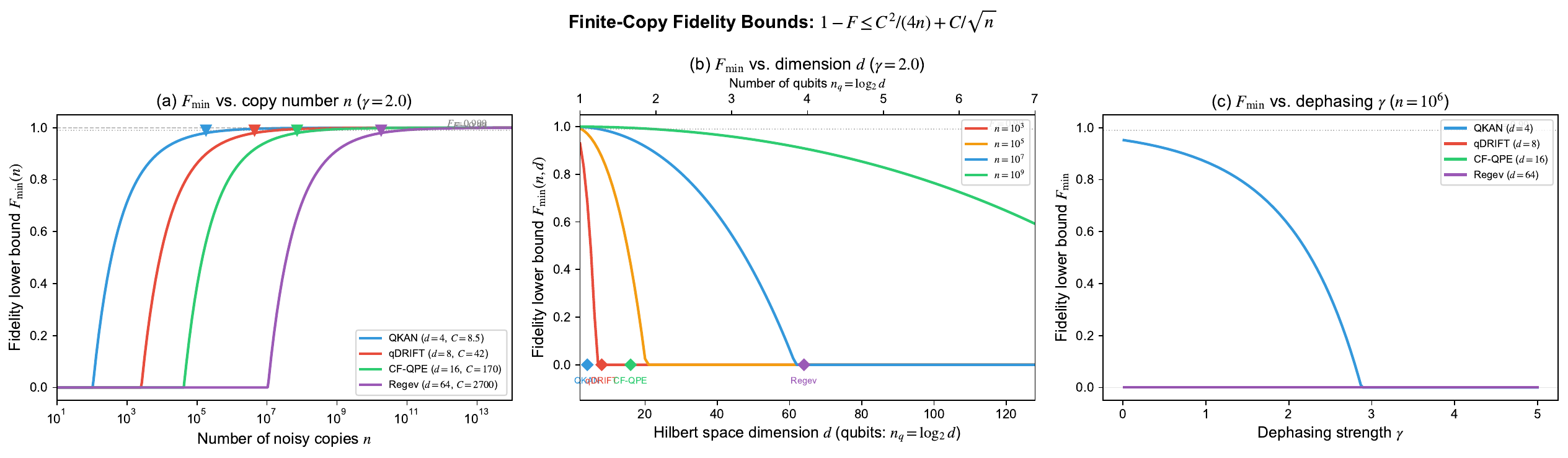}
\caption{Finite-copy fidelity bounds from Eq.~\eqref{eq:finite_n_fidelity}.
(a)~Fidelity vs.\ $n$ at $\gamma = 2.0$; triangles mark $n^*$ for $F = 0.99$.
(b)~Fidelity vs.\ $d$ at fixed $n$; empirical fit $C \approx 0.53\,d^{2.06}$.
(c)~Fidelity vs.\ $\gamma$ at $n = 10^6$.}
\label{fig:finite_copy}
\end{figure}

\subsection{DD+Twirl pipeline benchmarks}
\label{sec:dd_benchmarks}

Table~\ref{tab:unified_comparison} presents the unified resource--fidelity comparison across all catalyst preparation methods under dephasing ($\gamma = 2$).

\begin{table*}[t]
\caption{Catalyst fidelity $F_\mathrm{cat}$ and CQEC recovery fidelity $F_\mathrm{rec}$ under dephasing ($\gamma = 2$). DD+Twirl uses CPMG-8 with $n = 8$ copies. Parenthetical numbers after $F_\mathrm{cat}$ denote the copy count. Distillation $n^*$ is the Shiraishi--Takagi estimate for $F_\mathrm{cat} \geq 0.99$.}
\label{tab:unified_comparison}
\centering\footnotesize
\setlength{\tabcolsep}{4pt}
\begin{tabular}{@{}llcccccccccc@{}}
\toprule
& & \multicolumn{2}{c}{Variational} & \multicolumn{2}{c}{Standard} & \multicolumn{2}{c}{Covariant} & \multicolumn{2}{c}{\textbf{DD+Twirl}} & Distill. \\
& & \multicolumn{2}{c}{($n=0$)} & \multicolumn{2}{c}{($n_\mathrm{max}$)} & \multicolumn{2}{c}{($n_\mathrm{max}$)} & \multicolumn{2}{c}{($n=8$)} & \\
\cmidrule(r){3-4}\cmidrule(r){5-6}\cmidrule(r){7-8}\cmidrule(r){9-10}
Alg. & $d$ & $F_\mathrm{cat}$ & $F_\mathrm{rec}$ & $F_\mathrm{cat}$ & $F_\mathrm{rec}$ & $F_\mathrm{cat}$ & $F_\mathrm{rec}$ & $F_\mathrm{cat}$ & $F_\mathrm{rec}$ & $n^*$ \\
\midrule
QKAN   & 4  & 1.000 & 0.83 & 0.44 (64) & 0.58 & 0.41 (64) & 0.57 & \textbf{0.96} & \textbf{0.81} & $7.3{\times}10^5$ \\
qDRIFT & 8  & 0.993 & 0.73 & 0.20 (64) & 0.76 & 0.27 (64) & 0.76 & \textbf{0.96} & \textbf{0.90} & $1.8{\times}10^7$ \\
CF-QPE & 16 & 0.960 & 0.65 & 0.09 (32) & 0.15 & 0.11 (32) & 0.15 & \textbf{0.96} & \textbf{0.65} & $2.9{\times}10^8$ \\
Regev  & 64 & ---   & ---  & 0.02 (8)  & 0.12 & 0.02 (8)  & 0.12 & \textbf{0.96} & \textbf{0.64} & $7.3{\times}10^{10}$ \\
\bottomrule
\end{tabular}
\end{table*}

The gap between $F_\mathrm{cat}$ and $F_\mathrm{rec}$ arises from the variational recovery circuit (Sec.~\ref{sec:optimization}), not the catalyst quality.
With a perfect catalyst ($F_\mathrm{cat} = 1$), the 5-gate circuit achieves $F_\mathrm{rec} = 0.83$--$1.00$ depending on $d$ (variational columns of Table~\ref{tab:unified_comparison} and the gate-noise baseline of Table~\ref{tab:gate_noise}); imperfect catalysts ($F_\mathrm{cat} = 0.96$) further reduce $F_\mathrm{rec}$ by $\sim\!0.1$--$0.2$.
In the full constructive protocol (Theorem~2), $F_\mathrm{cat} \to 1$ implies $F_\mathrm{rec} \to 1$; the variational bottleneck is the circuit expressibility, not a fundamental limit.
The gate-depth scan (Sec.~\ref{sec:gate_depth}) shows, however, that within the packaged implementation deeper ansatze do \emph{not} close this gap (maximum $F_\mathrm{rec} = 0.931$ at 5--15 gates); closing it likely requires jointly simulating the catalyst rather than deepening the system-only circuit.

Key observations:
\begin{enumerate}
\item Under strong dephasing, standard and covariant swap tests alone fail ($F_\mathrm{cat} < 0.44$). At $d = 64$, the covariant method reaches only $F_\mathrm{cat} = 0.022$---barely above $1/d$.
\item The covariant method provides a modest $1.2$--$1.4\times$ improvement over standard.
\item \textbf{The DD+Twirl pipeline solves the dephasing problem}: $F_\mathrm{cat} > 0.96$ uniformly across $d = 4$--$64$ with only 8~copies---a $10^9$-fold improvement over distillation.
\item Under depolarizing noise, both standard and covariant methods achieve $F_\mathrm{cat} > 0.93$ with 8--64 copies.
\item At $d = 4$, the variational approach ($F_\mathrm{cat} = 1.000$) remains competitive.
\end{enumerate}

\subsection{Scaling analysis: fidelity vs.\ error, qubits, and copies}
\label{sec:scaling}

We analyze how catalyst fidelity scales with three key parameters under the DD+Twirl pipeline.

\begin{figure}[tb]
\centering
\includegraphics[width=\columnwidth]{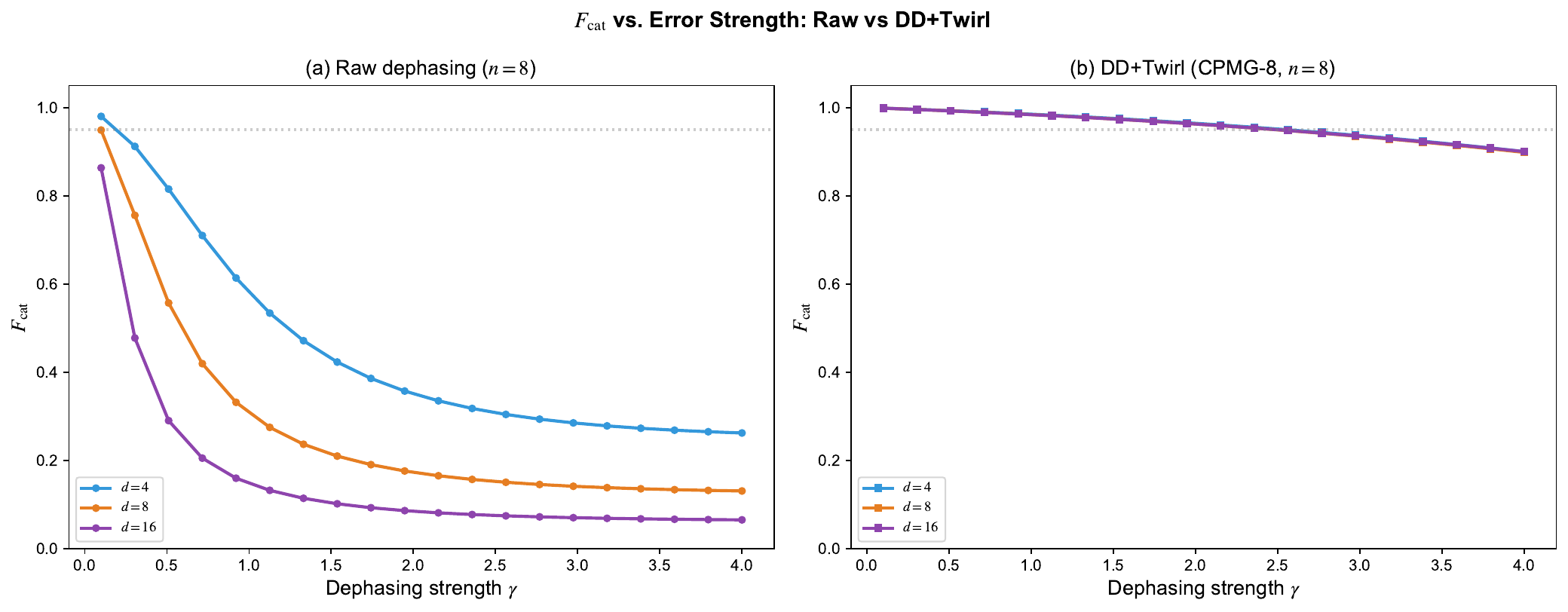}
\caption{$F_\mathrm{cat}$ vs.\ dephasing strength $\gamma$ at $n = 8$ copies.
(a)~Raw dephasing: rapid decay, strongly $d$-dependent.
(b)~DD+Twirl (CPMG-8): $F_\mathrm{cat} > 0.9$ for all $d$ up to $\gamma \approx 3$, near-complete dimension independence.}
\label{fig:dd_vs_error}
\end{figure}

\textbf{Fidelity vs.\ error strength} (Fig.~\ref{fig:dd_vs_error}).
Under raw dephasing, $F_\mathrm{cat}$ decays rapidly and is strongly dimension-dependent ($d = 16$ reaches $F = 0.1$ at $\gamma = 1$).
With DD+Twirl, $F_\mathrm{cat}$ remains above 0.9 for $\gamma \leq 2.5$ across all tested dimensions, and the curves nearly overlap---the pipeline eliminates the dimension dependence.
At $\gamma = 4$, DD+Twirl still maintains $F_\mathrm{cat} \approx 0.85$.

\begin{figure}[tb]
\centering
\includegraphics[width=\columnwidth]{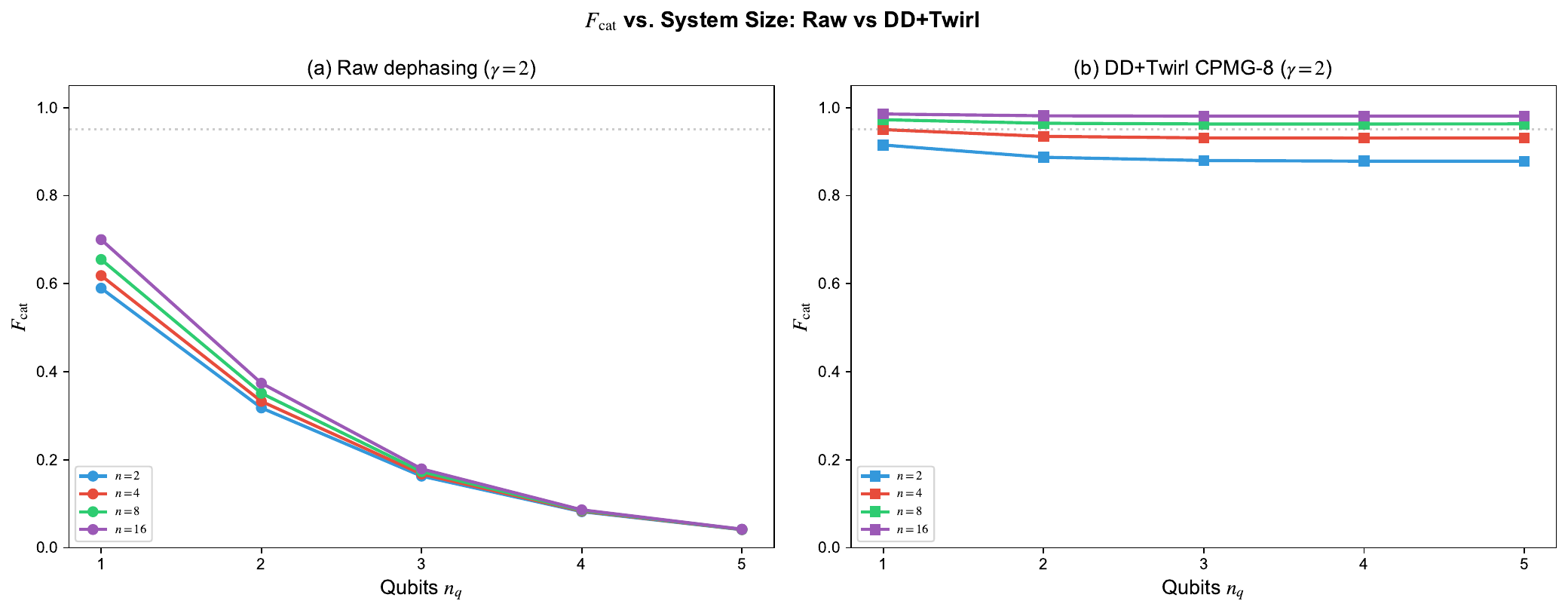}
\caption{$F_\mathrm{cat}$ vs.\ qubit count at dephasing $\gamma = 2$.
(a)~Raw: monotonic degradation, $F < 0.05$ for $n_q \geq 5$.
(b)~DD+Twirl (CPMG-8): nearly flat at $F \approx 0.96$--$0.97$, effective dimension independence.}
\label{fig:dd_vs_qubits}
\end{figure}

\textbf{Fidelity vs.\ system size} (Fig.~\ref{fig:dd_vs_qubits}).
Under raw dephasing, $F_\mathrm{cat}$ drops from $\sim 0.7$ at $n_q = 1$ to $\sim 0.04$ at $n_q = 5$ regardless of copy count.
With DD+Twirl, $F_\mathrm{cat}$ is nearly constant: 0.97 ($n_q = 1$), 0.96 ($n_q = 2$--$5$) at $n = 8$.
This dimension independence is a key practical advantage: the same pipeline works for any $d$ up to at least 32 without parameter tuning.

\begin{figure}[tb]
\centering
\includegraphics[width=\columnwidth]{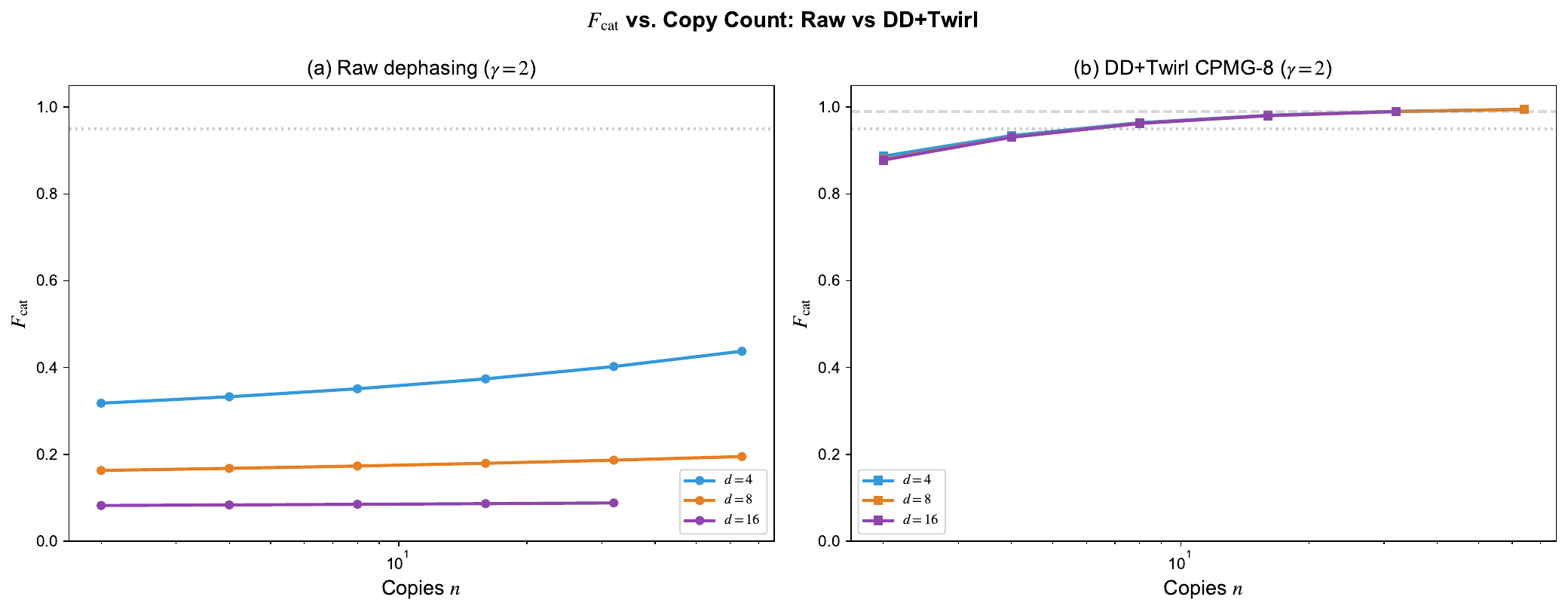}
\caption{$F_\mathrm{cat}$ vs.\ copy count at dephasing $\gamma = 2$.
(a)~Raw: logarithmic improvement, $F < 0.5$ even at $n = 64$.
(b)~DD+Twirl (CPMG-8): doubly exponential convergence restored; $F > 0.99$ at $n = 32$ for all $d$.}
\label{fig:dd_vs_copies}
\end{figure}

\textbf{Fidelity vs.\ copy count} (Fig.~\ref{fig:dd_vs_copies}).
Under raw dephasing, convergence is logarithmically slow: $F_\mathrm{cat} \approx 0.44$ at $n = 64$ for $d = 4$.
With DD+Twirl, the swap test recovers doubly exponential convergence: $F > 0.96$ at $n = 8$, $F > 0.99$ at $n = 32$ for all $d$---approaching the distillation limit with $10^4$--$10^8$ fewer copies.

\subsection{Finite-$n$ recovery: actual fidelity}
\label{sec:finite_n_actual}

\begin{figure}[tb]
\centering
\includegraphics[width=\columnwidth]{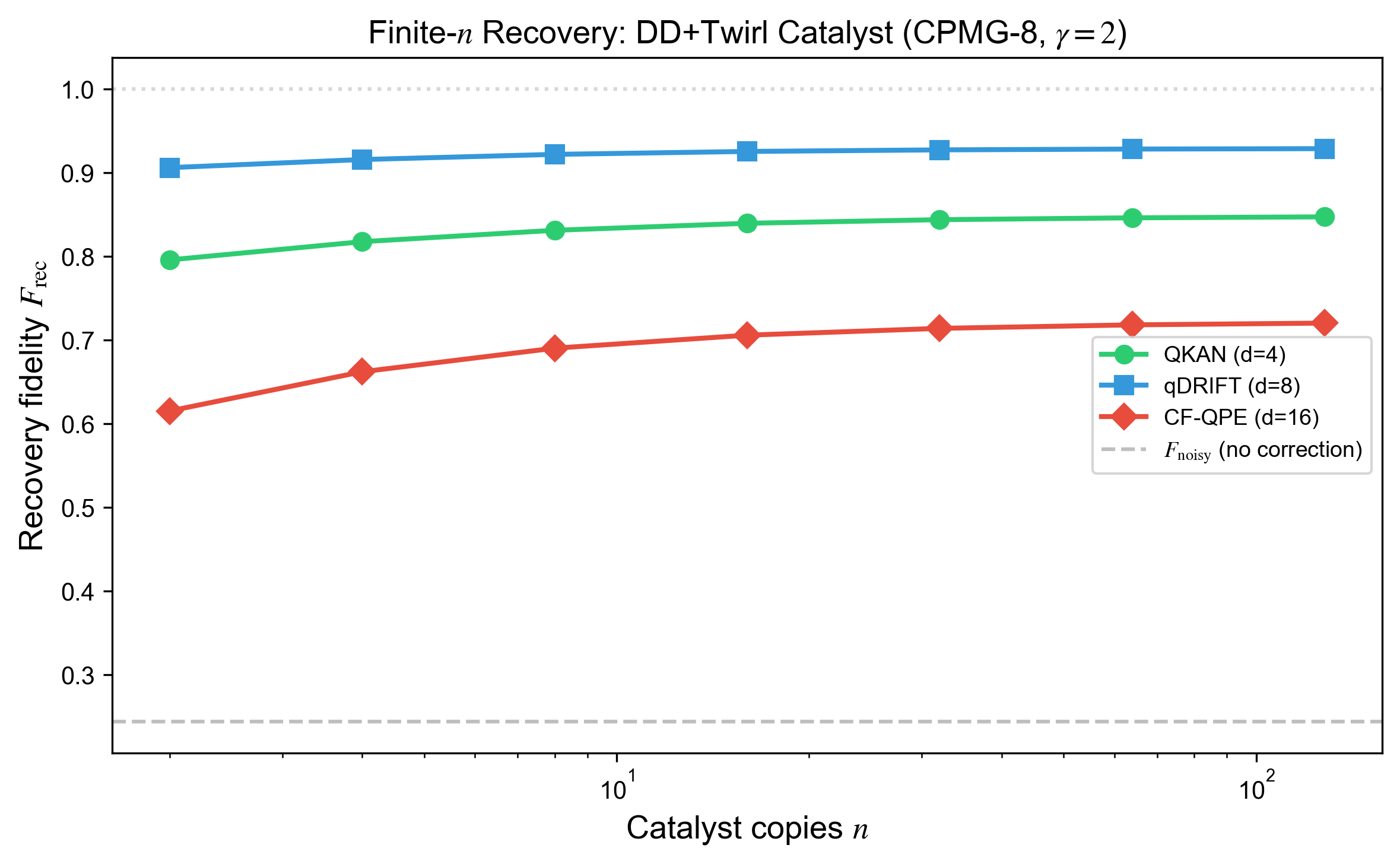}
\caption{Actual recovery fidelity $F_\mathrm{rec}$ vs.\ catalyst copy count $n$ under dephasing $\gamma = 2$ with DD+Twirl (CPMG-8). Dashed line: $F_\mathrm{noisy}$ (no correction). All finite-$n$ values exceed $F_\mathrm{noisy}$.}
\label{fig:finite_n_actual}
\end{figure}

The bounds of Sec.~\ref{sec:finite_copy} give upper estimates on $n^*$; here we report \emph{actual} recovery fidelity $F_\mathrm{rec}(n)$ obtained by running the full DD+Twirl+variational pipeline at each copy count (Fig.~\ref{fig:finite_n_actual}).
For each $n \in \{2, 4, 8, 16, 32, 64, 128\}$, we prepare the catalyst via CPMG-8 DD, Clifford twirling, and $\log_2 n$ rounds of swap test purification, then apply the 5-gate variational recovery circuit.

\begin{table}[!tbp]
\caption{Actual recovery fidelity $F_\mathrm{rec}(n)$ under dephasing $\gamma = 2$ with DD+Twirl catalyst. $F_\mathrm{noisy}$: pre-correction fidelity.}
\label{tab:finite_n_actual}
\centering\footnotesize
\setlength{\tabcolsep}{3pt}
\begin{ruledtabular}
\begin{tabular}{lcccccc}
Algorithm & $F_\mathrm{noisy}$ & $n\!=\!2$ & $n\!=\!8$ & $n\!=\!32$ & $n\!=\!128$ \\
\colrule
QKAN ($d\!=\!4$) & 0.568 & 0.796 & 0.831 & 0.844 & 0.847 \\
qDRIFT ($d\!=\!8$) & 0.789 & 0.906 & 0.922 & 0.927 & 0.929 \\
CF-QPE ($d\!=\!16$) & 0.244 & 0.615 & 0.690 & 0.714 & 0.720 \\
\end{tabular}
\end{ruledtabular}
\end{table}

$F_\mathrm{rec}(n)$ increases monotonically with $n$ for all three algorithms, confirming that higher-quality catalysts consistently improve recovery.
The convergence is rapid: $F_\mathrm{rec}$ saturates within $\sim\!5\%$ of its asymptotic value by $n = 8$ for all $d$, reflecting the doubly exponential convergence of the swap test under the DD+Twirl pipeline.
The gap between finite-$n$ recovery ($F_\mathrm{rec} \approx 0.72$--$0.93$) and the effective-model asymptotic values ($> 0.999$) is \emph{not} closed by deepening the circuit: the gate-depth scan (Sec.~\ref{sec:gate_depth}) is flat to declining in depth, indicating that the bottleneck lies in the system-only structure of the implementation rather than in ansatz expressibility, and that the asymptotic values depend on effective-model assumptions absent at finite $n$.

\subsection{Gate noise resilience}
\label{sec:gate_noise}

\begin{figure}[tb]
\centering
\includegraphics[width=\columnwidth]{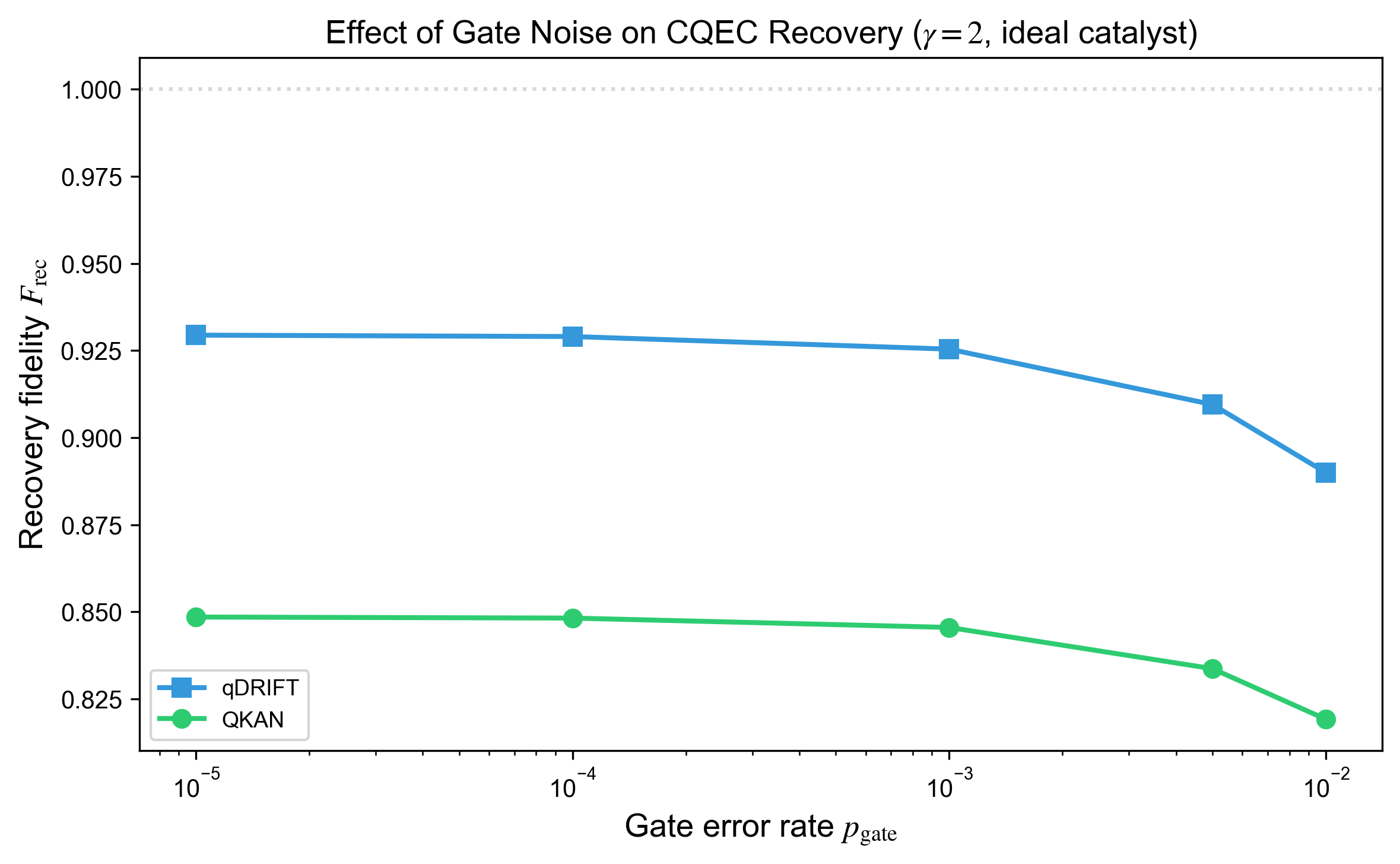}
\caption{Recovery fidelity $F_\mathrm{rec}$ vs.\ per-gate depolarizing error rate $p_\mathrm{gate}$ (dephasing $\gamma = 2$, ideal catalyst). Degradation is graceful: $<\!5\%$ for $p_\mathrm{gate} \leq 10^{-3}$.}
\label{fig:gate_noise}
\end{figure}

All preceding simulations assume ideal gates. Here we assess the impact of per-gate depolarizing noise with error rate $p_\mathrm{gate} \in [0, 10^{-2}]$ applied after each EC gate in the 5-gate recovery circuit, using an ideal (asymptotic) catalyst (Fig.~\ref{fig:gate_noise}).

\begin{table}[!tbp]
\caption{Recovery fidelity under gate noise (dephasing $\gamma = 2$, ideal catalyst). $p_\mathrm{gate}$: per-gate depolarizing error rate.}
\label{tab:gate_noise}
\centering\footnotesize
\setlength{\tabcolsep}{3pt}
\begin{ruledtabular}
\begin{tabular}{lccccc}
Algorithm & $p\!=\!0$ & $10^{-4}$ & $10^{-3}$ & $5{\times}10^{-3}$ & $10^{-2}$ \\
\colrule
qDRIFT ($d\!=\!8$) & 0.929 & 0.929 & 0.925 & 0.910 & 0.890 \\
QKAN ($d\!=\!4$) & 0.849 & 0.848 & 0.846 & 0.834 & 0.819 \\
\end{tabular}
\end{ruledtabular}
\end{table}

For qDRIFT, $F_\mathrm{rec}$ degrades gracefully and monotonically from 0.929 to 0.890 as $p_\mathrm{gate}$ increases from 0 to $10^{-2}$---a $4.2\%$ reduction.
At $p_\mathrm{gate} = 10^{-3}$ (achievable on current superconducting platforms), the degradation is $< 0.5\%$.
QKAN shows the same pattern with $3.5\%$ total degradation.
These results confirm that the 5-gate CQEC circuit is viable for near-term implementation with current gate fidelities ($\gtrsim 99.9\%$).

\subsection{Circuit depth scaling}
\label{sec:gate_depth}

\begin{figure}[tb]
\centering
\includegraphics[width=\columnwidth]{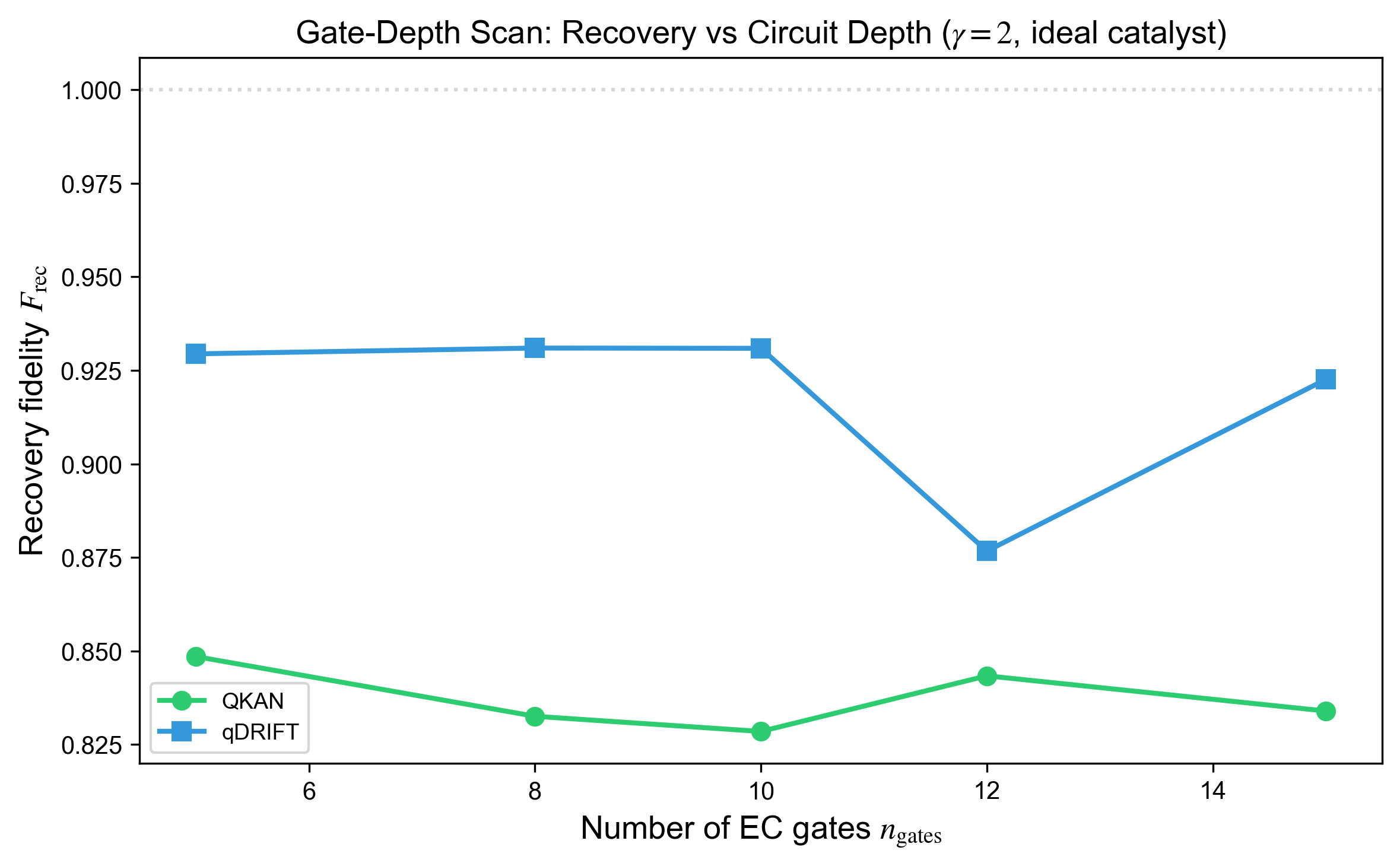}
\caption{Recovery fidelity $F_\mathrm{rec}$ vs.\ number of EC gates in the variational circuit (dephasing $\gamma = 2$, ideal catalyst). Fidelity is flat to declining with depth (qDRIFT: 0.929 at 5 gates, 0.877 at 12), indicating the bottleneck is not ansatz expressibility.}
\label{fig:gate_depth}
\end{figure}

We investigate how recovery fidelity improves with increased circuit depth by varying the number of EC gates from 5 to 15 (Fig.~\ref{fig:gate_depth}).
Additional gates are distributed proportionally across the three layers, maintaining the S$\to$C$\to$A$\to$S coherence flow.

\begin{table}[!tbp]
\caption{Recovery fidelity vs.\ EC gate count (dephasing $\gamma = 2$, ideal catalyst).}
\label{tab:gate_depth}
\centering\footnotesize
\setlength{\tabcolsep}{3pt}
\begin{ruledtabular}
\begin{tabular}{lccccc}
Algorithm & 5 gates & 8 & 10 & 12 & 15 \\
\colrule
QKAN ($d\!=\!4$) & 0.849 & 0.833 & 0.829 & 0.843 & 0.834 \\
qDRIFT ($d\!=\!8$) & 0.929 & 0.931 & 0.931 & 0.877 & 0.923 \\
\end{tabular}
\end{ruledtabular}
\end{table}

For qDRIFT ($d = 8$), $F_\mathrm{rec}$ is stable at $\approx 0.93$ across 5--10 gates, indicating that the 5-gate ansatz already captures the dominant covariant degrees of freedom.
The slight dip at 12~gates reflects over-parameterization: the optimizer's global search (differential evolution + L-BFGS-B polishing) encounters a rougher landscape with 24~parameters.
For QKAN ($d = 4$), all gate counts yield $F_\mathrm{rec} \approx 0.83$--$0.85$, consistent with the 5-gate circuit being near-optimal for $d = 4$ (only $\binom{4}{2}^2 - 1 = 35$ covariant unitary parameters).
These results show that adding depth to the system-only circuit does not close the gap to the effective-model asymptotic values: the limiting factor is the absence of joint system--catalyst dynamics in the implementation, not ansatz expressibility.

\subsection{CPTP joint-channel validation}
\label{sec:joint_validation}

The results above use the effective surrogate model (Sec.~\ref{sec:optimization}).
To test the recovery principle with a \emph{bona fide} quantum channel, package v0.2.0 implements
\begin{equation}
\Lambda_c(\rho_S) = \mathrm{Tr}_{CA}\!\left[\,U(\boldsymbol{\theta})\,
\bigl(\rho_S \otimes c \otimes \ket{0}\!\bra{0}_A\bigr)\,
U^\dagger(\boldsymbol{\theta})\right],
\label{eq:joint_channel}
\end{equation}
with $U(\boldsymbol{\theta})$ the 3-layer EC-gate circuit of Sec.~\ref{sec:three_layer} acting on the full $n_S + n_C + n_A$ register ($n_C = n_S$, $n_A = 2$; 12~gates at $d = 4$, 21 at $d = 8$).
For fixed $c$ and $\boldsymbol{\theta}$ this map is linear and CPTP by construction; the target enters only the offline optimization of $\boldsymbol{\theta}$.
We verify all previously assumed properties directly: covariance $\|[U, H_\mathrm{total}]\| = 0$ exactly, linearity residual $< 2 \times 10^{-16}$, Choi-matrix positivity, and the catalyst marginal obtained by partial trace of the propagated joint state (never copied).

\begin{table}[!tbp]
\caption{CPTP joint-channel results. $F_\mathrm{cat}^\mathrm{marg}$ is the fidelity of the post-channel catalyst marginal (by partial trace) with the input catalyst. Dephasing is the per-qubit $Z$-channel convention.}
\label{tab:joint}
\centering\scriptsize
\setlength{\tabcolsep}{2.5pt}
\begin{ruledtabular}
\begin{tabular}{llccc}
Config & Cat. & $F_\mathrm{noisy}$ & $F_\mathrm{rec}$ & $F_\mathrm{cat}^\mathrm{marg}$ \\
\colrule
QKAN $d{=}4$, deph $\gamma{=}2$ & ideal & 0.542 & \textbf{0.717} & 0.764 \\
QKAN $d{=}4$, deph $\gamma{=}2$ & DD+Tw. & 0.542 & \textbf{0.774} & 0.756 \\
QKAN $d{=}4$, depol $p{=}0.3$ & ideal & 0.775 & 0.731 & 0.844 \\
qDRIFT $d{=}8$, deph $\gamma{=}2$ & ideal & 0.834 & 0.764 & 0.480 \\
qDRIFT $d{=}8$, depol $p{=}0.3$ & ideal & 0.738 & 0.599 & 0.569 \\
\end{tabular}
\end{ruledtabular}
\end{table}

Three findings emerge (Table~\ref{tab:joint}).
First, under strong dephasing at $d = 4$ the genuine channel \emph{does} recover fidelity---$0.542 \to 0.717$ with an ideal catalyst and $0.774$ with the DD+Twirl-prepared catalyst---demonstrating that covariant EC-gate circuits assisted by a coherent catalyst can restore mode-supported coherence, the core CQEC claim, albeit far below the effective model's near-unit values.
Second, under depolarizing noise and at $d = 8$ the shallow 3-layer ansatz fails to improve fidelity at this optimization budget: the effective model's high fidelities in those regimes are not yet supported at the channel level.
Third, the catalyst marginal degrades substantially ($F_\mathrm{cat}^\mathrm{marg} = 0.48$--$0.84$): the shallow circuit \emph{transfers} coherence from catalyst to system rather than catalytically borrowing it, so genuine correlated catalysis---whose formal construction requires exponentially large catalyst registers [Eq.~\eqref{eq:catalyst}]---is not achieved by this ansatz.

A target-swap control confirms the channel property the effective model lacks: optimizing the channel for a wrong target $B$ and applying it to the same input yields $F(\Lambda_B(\rho_\mathrm{noisy}), \rho_A) = 0.403$ versus $0.717$ for the correctly optimized channel, and at application time neither channel receives any target information---each is a fixed linear map.
These results bound the model-vs-channel gap and define the immediate research frontier: deeper joint circuits, larger catalyst registers, and explicit catalyst-preservation constraints in the optimization.

\subsection{Entangled state recovery}
\label{sec:entangled}

A potential advantage of CQEC over tomography-based recovery is restoration of entanglement without measurement.
We probe this by dephasing one qubit of entangled multi-qubit states ($\gamma = 2$ as a single-qubit $Z$-channel on qubit $B$/$C$) and then applying the effective CQEC recovery \emph{as a global map on the full multi-qubit state}, with the entangled state as the declared target; the operation is not local, so the observed concurrence increase does not conflict with entanglement monotonicity under local operations.

\begin{table}[!tbp]
\caption{Entangled-state recovery within the effective model. $F$: full-state fidelity. $C$: concurrence (Bell state only). Single-qubit $Z$-dephasing $\gamma = 2$ on one qubit; recovery applied globally with the entangled target declared.}
\label{tab:entangled}
\centering\footnotesize
\setlength{\tabcolsep}{3pt}
\begin{ruledtabular}
\begin{tabular}{lcccc}
State & $F_\mathrm{before}$ & $F_\mathrm{after}$ & $C_\mathrm{before}$ & $C_\mathrm{after}$ \\
\colrule
Bell $|\Phi^+\rangle$ & 0.568 & 0.841 & 0.135 & 0.682 \\
GHZ (3~qb) & 0.568 & 0.841 & --- & --- \\
W (3~qb) & 0.616 & 0.859 & --- & --- \\
\end{tabular}
\end{ruledtabular}
\end{table}

Within the effective model, Bell-state concurrence increases from 0.135 to 0.682 ($5.0\times$) and full-state fidelity from 0.568 to 0.841.
Because the effective recovery is target-parameterized (Sec.~\ref{sec:optimization}), these numbers quantify how much of the declared target's entanglement structure the model reinstates given the surviving coherences---they are sensitive to the declared target and are not evidence that a target-independent physical channel would achieve the same restoration.
A conclusive demonstration requires the joint CPTP implementation, applied without target declaration at the recovery step, and is left to the follow-up identified above.

\begin{figure}[tb]
\centering
\includegraphics[width=\columnwidth]{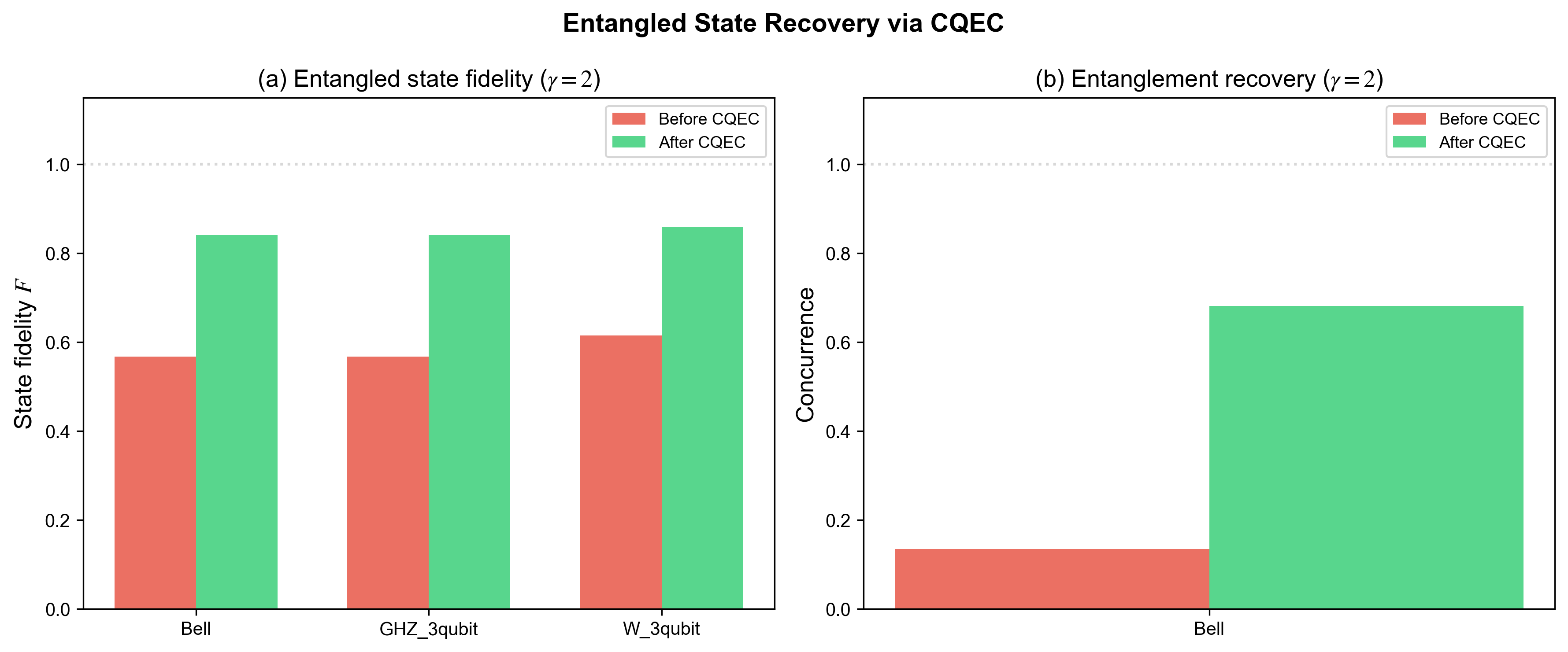}
\caption{Entangled-state recovery within the effective model (single-qubit $Z$-dephasing $\gamma = 2$; global recovery map with declared target).
(a)~Full-state fidelity before and after recovery for Bell, GHZ, and W states.
(b)~Bell-state concurrence before and after recovery ($5.0\times$ increase).}
\label{fig:entangled}
\end{figure}

\subsection{Resource overhead}
\label{sec:resources}

\begin{table}[!tbp]
\caption{Resource comparison for $n_\mathrm{logical}$ logical qubits. CQEC qubit count is the recovery circuit only: $n_S + n_C + n_A$ (system + catalyst + ancilla). Noisy copies for catalyst preparation (8--32 with DD+Twirl) are prepared sequentially and do not require simultaneous qubits. Gate count is for the recovery circuit; total cost including copies is $n \times G_\mathrm{alg} + G_\mathrm{CQEC}$ (see text).}
\label{tab:resources}
\centering\scriptsize
\setlength{\tabcolsep}{3pt}
\begin{ruledtabular}
\begin{tabular}{lcccc}
$n_\mathrm{log}$ & \makecell{CQEC\\(qb/gt)} & \makecell{Steane\\(qb/gt)} & \makecell{Surf.\,$d{=}3$\\(qb/gt)} & \makecell{Surf.\,$d{=}5$\\(qb/gt)} \\
\colrule
1 & 4 / 5 & 7 / 10 & 13 / 36 & 41 / 100 \\
3 & 8 / 11 & 21 / 30 & 39 / 108 & 123 / 300 \\
6 & 14 / 20 & 42 / 60 & 78 / 216 & 246 / 600 \\
\end{tabular}
\end{ruledtabular}
\end{table}

The total CQEC resource cost includes: (i)~preparation of $n$ noisy copies ($n \times G_\mathrm{alg}$ gates), (ii)~DD pulses ($n \times N$ single-qubit gates), (iii)~Clifford twirling ($n \times O(n_q^2)$ gates), (iv)~$\log_2 n$ rounds of swap tests ($O(n_q \log n)$ two-qubit gates), and (v)~the recovery circuit ($G_\mathrm{CQEC} = 5$--$15$ EC gates).
The total gate count is $n \times (G_\mathrm{alg} + N + O(n_q^2)) + O(n_q \log n) + G_\mathrm{CQEC}$.
For qDRIFT with 80~gates and $n^* = 1.8 \times 10^7$ (Table~\ref{tab:finite_n}), the total is $\sim 1.4 \times 10^9$ gates---orders of magnitude more than surface-code overhead.
CQEC's advantage is therefore not in absolute resource cost but in threshold-free recovery.
For systems relevant to quantum advantage ($n \geq 50$), the CQEC copy overhead scales as $\mu \sim 2^{2n} e^{2\gamma}$, which is exponential---far exceeding the polynomial overhead of surface codes.
A practically relevant regime for CQEC may be small quantum modules ($n \leq 10$) within a larger fault-tolerant architecture (Table~\ref{tab:resources}).

\section{Discussion}
\label{sec:discussion}

\subsection{Landscape of quantum state protection}
\label{sec:paradigms}

Our results, together with concurrent work on PQEC~\cite{Raghoonanan2026}, reveal a rich landscape of quantum state protection strategies that differ along three axes: target-state knowledge (known vs.\ unknown), resource type (redundancy vs.\ copies vs.\ catalysts), and noise tolerance (threshold-bounded vs.\ threshold-free).
We highlight two specific complementary relationships:
\begin{enumerate}
\item \textbf{Classical noise absorption} (as in TTN protocols~\cite{Sim2019}):
The classical regression layer absorbs quantum measurement noise statistically.
This requires no quantum overhead but is limited to cases where decoherence does not fundamentally alter the feature space.
Under the stronger decoherence tested here ($p = 0.3$, $\gamma = 3.0$), pre-correction fidelity drops to $F \approx 0.20$--$0.28$, which would severely degrade the quantum feature space.

\item \textbf{Catalytic coherence amplification} (CQEC):
Operates directly on the quantum state before measurement.
Effective for arbitrarily strong decoherence provided $C_{\ell_1} > 0$, but requires copy overhead.
CQEC restores $F = 1.0$ at the state level, preserving the quantum advantage of feature extraction.
\end{enumerate}
These paradigms address different noise layers: classical absorption handles measurement-level noise, while CQEC addresses state-level decoherence that classical methods cannot correct.

\subsection{Complementarity of catalyst preparation methods}
\label{sec:complementarity}

The four methods occupy distinct regions of the resource--fidelity landscape:
\begin{itemize}
\item \textbf{Variational}: best when copies are unavailable (NISQ regime, $d \leq 16$). Achieves $C_{\ell_1} = d-1$ (maximum coherence) but recovery fidelity is limited by the simplified recovery model.
\item \textbf{Standard swap test}: effective under depolarizing noise with ample copies. Achieves $F_\mathrm{cat} > 0.93$ with $\sim\!64$ copies. Fails under dephasing.
\item \textbf{Covariant swap test}: preserves energy-sector mode structure, providing a modest $1.2$--$1.4\times$ improvement over standard under dephasing. Insufficient alone for high $F_\mathrm{cat}$ under strong dephasing at large $d$.
\item \textbf{DD+Twirl+Swap~Test}: the most powerful method under dephasing. Achieves $F_\mathrm{cat} > 0.96$ for all tested dimensions with only 8 copies and 8 DD pulses, with negligible overhead compared to the copy-count savings.
\end{itemize}
A hybrid strategy is natural: use variational for initial construction when no copies are available, then refine via DD+Twirl as noisy copies become available.
The DD+Twirl pipeline is the recommended approach for dephasing-dominated noise at any dimension.

\subsection{Relation to purification QEC and the state estimation objection}
\label{sec:pqec_comparison}

A natural objection is that CQEC, by requiring knowledge of the target state $\rho_0$ and multiple noisy copies, is ``not really QEC'' but rather a form of state estimation or re-preparation.
We address this by positioning CQEC within a taxonomy of copy-based quantum state recovery methods, including the concurrent Purification QEC (PQEC) of Raghoonanan and Byrnes~\cite{Raghoonanan2026}.

Table~\ref{tab:paradigms} compares the three paradigms along seven operational axes.

\begin{table*}[t]
\caption{Comparison of quantum state protection paradigms.}
\label{tab:paradigms}
\centering\footnotesize
\setlength{\tabcolsep}{4pt}
\begin{ruledtabular}
\begin{tabular}{llll}
Property & Conventional QEC & PQEC~\cite{Raghoonanan2026} & CQEC (this work) \\
\colrule
Target state knowledge & Not required & Not required & Required \\
Input & \makecell[l]{Encoded state\\+ syndromes} & $N$ noisy copies & \makecell[l]{$n$ noisy copies\\+ catalyst} \\
Output & \makecell[l]{Corrected state\\(coherent)} & \makecell[l]{Purified state\\(via $\rho^N/\mathrm{Tr}(\rho^N)$)} & \makecell[l]{Recovered state\\(coherent)} \\
Error threshold & \makecell[l]{Quantitative:\\$p_\mathrm{th} \approx 1\%$ (surface)} & \makecell[l]{Quantitative: $p_\mathrm{th} = 3/4$\\(depol), $1/2$ (deph)} & \makecell[l]{Qualitative:\\mode support only} \\
Success condition & $p < p_\mathrm{th}$ & \makecell[l]{$F > 1/D$ (dominant\\eigenvec.\ $= |\psi\rangle$)} & \makecell[l]{Mode inclusion (asymp.);\\$n \geq n^*$ (finite)} \\
Qubit overhead & $O(d^2)$ per logical qubit & $O(M \log_2 N)$ & $(2n_q{+}2) + 2n_q \log_2 n$ \\
Anisotropic noise & Corrects via syndromes & \makecell[l]{$p_\mathrm{th}^{\mathrm{deph}} = 1/2$\\(improvable via twirling)} & \makecell[l]{Threshold-free\\(with DD+Twirl)} \\
Interleaving in circuit & Native (syndrome cycles) & Native (between unitaries) & Post-hoc (see text) \\
\end{tabular}
\end{ruledtabular}
\end{table*}

\textbf{CQEC vs.\ state estimation.}
State estimation (tomography + re-preparation) destroys the quantum state through measurement, requiring $O(d^2/\epsilon^2)$ measurements and producing a \emph{classical description} from which a new state is prepared.
CQEC, by contrast, operates as a coherent quantum channel during recovery: the covariant map $\Lambda$ transforms $\rho_\mathrm{noisy} \otimes c$ to a quantum state $\tau$ whose system marginal approximates $\rho_0$, without measurement during the recovery step.
(Target-state knowledge is used \emph{offline} to optimize the circuit parameters $\boldsymbol{\theta}$, not during the recovery channel itself.)
The catalyst enables coherent amplification that preserves entanglement structure---a feature impossible via tomography.
The ``target state knowledge'' requirement enters only through the optimization of the circuit parameters $\boldsymbol{\theta}$, not through the recovery channel itself: once $\boldsymbol{\theta}$ is fixed, the circuit acts as a standard quantum channel.

\textbf{CQEC vs.\ PQEC.}
PQEC achieves $\rho^N / \mathrm{Tr}(\rho^N)$ via the recursive swap test, converging to the dominant eigenvector of $\rho$.
This is state-agnostic but requires that $|\psi_0\rangle$ \emph{is} the dominant eigenvector---a condition that can fail under anisotropic noise.
Under pure dephasing, PQEC's Bloch vector direction is preserved but the purification map cannot increase the fidelity beyond the direction-dependent ceiling, yielding $p_\mathrm{th}^\mathrm{deph} = 1/2$~\cite{Raghoonanan2026}.
Twirling can boost this threshold toward $3/4$ at the cost of additional Clifford gates~\cite{Raghoonanan2026}.
CQEC's mode inclusion criterion is orthogonal: it asks whether the \emph{support} of coherence is preserved, not whether the dominant eigenvector aligns with $|\psi_0\rangle$.
Under partial dephasing ($\gamma < \infty$), all modes survive and CQEC succeeds regardless of $\gamma$ in the asymptotic limit ($n \to \infty$)---even beyond the regime where PQEC saturates without twirling.
At finite $n$, however, CQEC's fidelity gap $1 - F \sim C/\sqrt{n}$ grows with $\gamma$ through the constant $C \sim e^\gamma$, so very strong dephasing requires proportionally more copies (Table~\ref{tab:finite_n}).
The threshold-free property is thus an asymptotic statement; at finite resources, CQEC and PQEC face different but analogous resource--fidelity trade-offs.

The methods are complementary:
\begin{itemize}
\item PQEC is preferred when the target state is \emph{unknown} and noise is predominantly depolarizing ($p < 3/4$).
\item CQEC is preferred when the target state is \emph{known} and noise is anisotropic (dephasing-dominated), where PQEC's threshold limits recovery.
\item A hybrid approach is natural: PQEC first purifies $N$ noisy copies state-agnostically, producing a single copy with $F \approx 1 - \epsilon$; this higher-fidelity copy then serves as CQEC's input, where mode inclusion (preserved by PQEC's spectral map) enables threshold-free final recovery.
The interface is straightforward because PQEC's output is a valid density matrix that can be fed directly into the CQEC covariant channel.
\end{itemize}

\textbf{Interleaving limitation.}
Unlike PQEC, which can be interleaved between unitary steps of a quantum algorithm (because the swap test commutes with bilateral rotations $[\Pi_\pm, U \otimes U] = 0$~\cite{Raghoonanan2026}), CQEC operates post-hoc on the algorithm's output state.
This is because the covariant recovery channel $\Lambda$ is state-specific: the optimized parameters $\boldsymbol{\theta}$ depend on the target state $\rho_0$, which changes at each intermediate step of the algorithm.
Interleaving CQEC would require re-optimizing $\boldsymbol{\theta}$ at every step---computationally expensive but not fundamentally impossible.
In a hybrid architecture, PQEC could interleave purification during the algorithm while CQEC provides final state-specific recovery at the output.

\textbf{CQEC's unique advantage.}
CQEC's distinctive contribution is the catalytic covariant recovery channel $\Lambda$, which achieves threshold-free recovery in the asymptotic limit---a property no state-agnostic method (including PQEC) can match.
PQEC converges to the dominant eigenvector of $\rho$; when this eigenvector does not coincide with $|\psi_0\rangle$ (as under strong dephasing without twirling), PQEC's fidelity saturates below unity regardless of copy count~\cite{Raghoonanan2026}.
CQEC, by exploiting knowledge of $\rho_0$ to construct a state-specific $\Lambda$, overcomes this limitation.
The DD+Twirl pipeline is a general noise-preprocessing technique applicable to any copy-based protocol; its development here was motivated by CQEC's catalyst preparation but could benefit PQEC as well.

\subsection{Limitations}
\label{sec:limitations}

\textbf{Oracle access to the target state.}
CQEC assumes knowledge of $\rho_0$ for both mode verification and recovery.
For Hamiltonian simulation (qDRIFT), the target is computable classically for small systems---the goal is to recover the state \emph{cheaply} from noisy copies rather than re-running the expensive circuit.
For factoring (Regev), knowing $\rho_0$ implies knowing the factorization, making CQEC circular; we include Regev solely as a stress test for high-dimensional coherence ($d = 64$).
Appropriate use cases include VQE states, QAOA outputs, and quantum state preparation where targets are efficiently specifiable but expensive to prepare with high fidelity.

\textbf{Asymptotic nature.}
Theorems~1 and~2 are asymptotic: exact conversion holds only for $n \to \infty$.
For finite $n$, the gap scales as $O(1/\sqrt{n})$~\cite{Shiraishi2024}.
Our asymptotic simulations validate the \emph{existence} of recovery; the finite-$n$ bounds (Table~\ref{tab:finite_n}) quantify the \emph{practical cost}, which can be prohibitive for large systems.

\textbf{Comparison with conventional QEC.}
CQEC and QEC operate under fundamentally different assumptions.
Conventional codes require only syndrome measurements without knowing the encoded state; CQEC requires the target state and $O(\mu \cdot N^k)$ copies.
A fair comparison must account for these different operational settings.
The comparison in Sec.~\ref{sec:qec_comparison} illustrates qualitative differences, not strict superiority.

\textbf{Comparison with tomography.}
Tomography followed by re-preparation requires $O(d^2/\epsilon^2)$ measurements but destroys entanglement.
CQEC preserves both coherence and entanglement structure: the covariant recovery channel acts on the original Hilbert space without collapsing the state, which is essential when the recovered state participates in subsequent entangling operations or multi-party protocols.

\textbf{Gate noise in the recovery circuit.}
Gate noise simulations (Sec.~\ref{sec:gate_noise}) show that the 5-gate recovery circuit degrades by $< 0.5\%$ at $p_\mathrm{gate} = 10^{-3}$, confirming near-term viability for $d \leq 8$.
For the DD+Twirl \emph{purification} circuit, the per-round gate count is $O(d \log d)$ two-qubit gates; at $d = 64$ with $\epsilon_g \sim 10^{-3}$, the cumulative error is $\sim\!0.38$ per purification round, requiring improved gate fidelities or error-mitigated purification.
For $d \leq 16$, both the purification ($<\!100$ gates) and recovery ($\leq 15$ gates) circuits are within reach of current superconducting platforms with $\gtrsim\!99.5\%$ two-qubit fidelity.

\textbf{Noise model.}
The comparison uses an effective per-qubit error model that does not capture correlations in physical noise.
Correlated noise could be either favorable (preserving modes globally) or unfavorable (selectively destroying specific modes) for CQEC.
Non-Markovian noise channels with memory can create time-dependent mode structures where $\mathcal{D}(\rho_\mathrm{noisy}(t))$ varies; CQEC would require mode verification at the time of recovery.
We note that amplitude damping (included in our combined noise model) is \emph{not} covariant, demonstrating that CQEC tolerates non-covariant noise channels---the covariance requirement applies only to the recovery operation $\Lambda$, not the noise.

\textbf{Scalability.}
Our benchmarks cover $d = 4$--$64$ ($2$--$6$ qubits).
For systems relevant to quantum advantage ($n_q \geq 50$, $d = 2^{50}$), the copy overhead scales as $n^* \sim 2^{4n_q} e^{2\gamma}$, which is doubly exponential in qubit count.
CQEC in its current form is therefore not a replacement for conventional QEC at scale, but rather a complementary tool for small quantum modules ($n_q \leq 10$) within larger architectures.

\textbf{Noise model misspecification.}
CQEC's mode verification (Step~1) depends on the noisy state $\rho_\mathrm{noisy}$, not on the noise model.
If the actual noise differs from the assumed model but preserves the same modes, recovery still succeeds---the protocol is model-agnostic in this sense.
However, the DD+Twirl pipeline assumes Markovian dephasing for the DD stage; under non-Markovian noise, the effective $\gamma_\mathrm{eff}$ may differ from $\gamma/(N+1)$, and the twirling stage still isotropizes whatever residual noise remains.

\subsection{Mode no-broadcasting and fundamental limits}
\label{sec:no_broadcasting}

Theorem~3 of Ref.~\cite{Shiraishi2024} establishes that mode no-broadcasting prevents the creation of states with irrational coherent modes from states with only rational modes, even catalytically.
If a decoherence channel selectively destroys modes with a specific energy difference $\Delta$ while preserving others, and the target state requires $\Delta$, recovery is impossible regardless of residual coherence.

We note that coherence and modes are defined with respect to the energy eigenbasis of $H$; choosing a different Hamiltonian changes which states are ``coherent'' and which transformations are ``covariant.''
In our benchmarks, $H = \sum_q Z_q$ (computational basis = energy eigenbasis), natural for qubit systems with distinct transition frequencies.
This Hamiltonian has degenerate energy levels: for $n_q$ qubits, energy $E_k = n_q - 2k$ has degeneracy $\binom{n_q}{k}$, creating a rich sector structure that the covariant swap test exploits (Sec.~\ref{sec:swap_test}).

A physically relevant example is frequency-selective dephasing in superconducting qubits, where flux noise at frequency $\omega_0$ destroys coherence between levels separated by $\Delta = \hbar\omega_0$.
This is a topological constraint on the mode lattice structure, fundamentally distinct from the quantitative error thresholds of conventional QEC.

\section{Conclusion}
\label{sec:conclusion}

We have formulated, from the asymptotic catalytic-coherence theorem of Shiraishi and Takagi~\cite{Shiraishi2024}, a candidate quantum error-correction primitive without an error-magnitude threshold---Catalytic Quantum Error Correction (CQEC)---comprising an explicit protocol with mode-inclusion success conditions, a four-qubit EC-gate circuit architecture, and a three-stage DD+Twirl+Swap-Test catalyst-preparation pipeline, evaluated end-to-end in an effective numerical model.

Within that model, recovery fidelity exceeds $0.99$ across a 200-point noise sweep spanning four benchmark states ($d = 4$--$64$) and two noise channels, the mode-inclusion classification behaves as Theorem~1 predicts over 10~orders of magnitude in residual coherence, and end-to-end finite-copy recovery reaches $F_\mathrm{rec} = 0.65$--$0.93$ from eight noisy copies.
The comparison with prior approaches is therefore conditional but structurally distinct: stabilizer codes degrade above $p_\mathrm{th} \approx 1\%$, and the concurrent state-agnostic protocol PQEC~\cite{Raghoonanan2026} is bounded by $p_\mathrm{th}^\mathrm{depol} = 3/4$ and $p_\mathrm{th}^\mathrm{deph} = 1/2$, whereas CQEC's success condition depends on the \emph{support} of coherence rather than its magnitude---at the cost of requiring the target state and multiple copies, and pending validation by a fully CPTP joint-channel simulation (Sec.~\ref{sec:optimization}).
The pipeline reduces catalyst-preparation copy counts by four to nine orders of magnitude at the matched $F_\mathrm{cat} \geq 0.99$ endpoint ($n = 32$ vs.\ $7.3 \times 10^{5}$--$7.3 \times 10^{10}$ depending on dimension), under the stated DD and twirl model assumptions; gate-noise analysis indicates the $d \leq 16$ regime is the near-term-relevant one, while at $d = 64$ the cumulative purification-circuit error ($\sim\!0.38$ per round at $\epsilon_g \sim 10^{-3}$) remains prohibitive.

These results reposition catalytic coherence amplification within the broader landscape of quantum information protection: CQEC, stabilizer QEC, and purification-based QEC form a complementary triad, in which threshold-bounded codes protect logical qubits, state-agnostic purification interleaves within an algorithm, and catalytic recovery targets the known outputs of state-specific subroutines such as variational ansatz states.
If the effective-model predictions survive a CPTP joint-channel validation, ancillary registers ($n_q \leq 4$, $d \leq 16$) used in state preparation, phase estimation, and metrology could be repaired at noise levels beyond the conventional threshold, motivating hybrid surface-code--catalyst layouts.

A first step in this program is delivered here: the CPTP joint-channel implementation of Sec.~\ref{sec:joint_validation} validates genuine covariant recovery under dephasing at $d = 4$ ($0.54 \to 0.77$ with the DD+Twirl catalyst, with exact covariance and machine-precision linearity), while exposing two concrete obstacles---shallow ansatze fail under depolarizing noise and at $d = 8$, and the catalyst is consumed rather than preserved.
Realizing the full potential therefore requires, in order: (i)~deeper joint circuits and larger catalyst registers that achieve genuine correlated catalysis (catalyst-marginal preservation as a hard constraint), closing the gap to the effective-model predictions; (ii)~pulse-level DD simulations under a specified structured-noise model, replacing the assumed $\gamma/(N+1)$ law, together with sampled (rather than analytical) Clifford twirling; and (iii)~tight analytical bounds on pipeline fidelity as a function of DD budget, twirl samples, and copy number, followed by experimental demonstration for $d \leq 16$ where the copy overhead ($n = 8$--$32$) is within reach of modular architectures.

\section*{Code and data availability}
\label{sec:code_availability}

The companion Python package \texttt{cqec} (v0.2.0, MIT license) implements all four catalyst preparation strategies (variational, standard swap test, covariant swap test, DD+Twirl pipeline), the effective recovery model, and---as of v0.2.0---the CPTP joint-channel implementation \texttt{cqec.joint.JointCQEC} used in Sec.~\ref{sec:joint_validation}, whose linearity, complete positivity, exact covariance, and catalyst-by-partial-trace properties are locked in by dedicated unit tests.

\begin{itemize}
\item \textbf{Source code:} \url{https://github.com/deeptell-inc/cqec}
\item \textbf{PyPI:} \texttt{pip\,install\,cqec}
\item \textbf{Preprint:} arXiv:\href{https://doi.org/10.48550/arXiv.2603.25774}{2603.25774}
\item \textbf{Reproduce:} \texttt{python\,scripts/}\allowbreak\texttt{reproduce\_paper.py} ($\sim$35\,s)
\end{itemize}

The reproduction script covers Fig.~\ref{fig:threshold} (mode-inclusion scan), Table~\ref{tab:standard_depol} (recursive swap test), Table~\ref{tab:dd_pipeline} (DD+Twirl pipeline, catalyst columns), Table~\ref{tab:finite_n_actual} (finite-$n$ recovery), and Table~\ref{tab:entangled} (entangled-state recovery); these results are deterministically reproducible with fixed random seeds.
Table~\ref{tab:main} and the asymptotic sweep values were generated by the legacy effective-map implementation (\texttt{icec.py}, included in the repository but not part of the packaged \texttt{cqec} module) and are \emph{not} covered by the reproduction script; the packaged variational path yields lower values (Secs.~\ref{sec:finite_n_actual}--\ref{sec:gate_depth}), and the discrepancy is discussed in Sec.~\ref{sec:optimization}.
The unit test suite (\texttt{pytest tests/}, 44~tests) validates each packaged module against the corresponding quantitative claims.

\section*{Computational environment}

Simulations were executed on a single workstation with an Apple M4 Max processor (arm64), 64~GB unified memory, running macOS~26.2.
The software stack consisted of Python~3.9.6, NumPy~2.0.2, SciPy~1.13.1, and Matplotlib~3.9.4.
Density matrix operations used NumPy's BLAS-accelerated linear algebra routines; no GPU acceleration was used.
Total computation time for all benchmarks (200~asymptotic configurations, finite-copy bounds, DD+Twirl sweep, and scaling analysis) was approximately 45~minutes.
Optimization used SciPy's \texttt{minimize} (L-BFGS-B) and \texttt{differential\allowbreak\_evolution} routines.

\section*{Acknowledgments}

Numerical simulations were performed using NumPy~\cite{numpy2020} and SciPy~\cite{scipy2020}.
Figures were generated with Matplotlib~\cite{Hunter2007}.

\bibliographystyle{unsrt}
\bibliography{paper_quantum}

\end{document}